\documentclass[journal=jacsat,manuscript=article]{achemso}
\usepackage{epsfig}
\usepackage{graphicx}
\usepackage{dcolumn}
\usepackage{amsthm,amsmath}
\usepackage{comment}
\usepackage[colorlinks]{hyperref}
\usepackage{xcolor}
\usepackage{longtable}
\usepackage{amsmath}

\author{Arup Chakraborty}
\affiliation{
Atomic, Molecular and Optical Physics Division, Physical Research Laboratory, Navrangpura, Ahmedabad 380009, India
and \\
Indian Institute of Technology Gandhinagar, Palaj, Gandhinagar 382355, India
}
\email{arupc794@gmail.com}
\author{Bijaya Kumar Sahoo}
\affiliation{
Atomic, Molecular and Optical Physics Division, Physical Research Laboratory, Navrangpura, Ahmedabad 380009, India
}
\email{bijaya@prl.res.in}

\title{Deciphering Core, Valence and Double-Core-Polarization Contributions to Parity Violating Amplitudes in $^{133}$Cs using Different Many-body Methods}

\begin{document}
\begin{abstract}
This work examines the accuracy of different many-body methods for the  calculations of parity violating electric dipole ($E1_{\text{PV}}$) amplitudes in atomic systems. In the last decade many different groups claim to achieve its accuracy below 0.5\%, for the $6s ~ ^2S_{1/2} \rightarrow 7s ~ ^2S_{1/2} $ transition in $^{133}$Cs atom. One of the major issues in these calculations is the opposite signs among the core correlation contribution from different works. To estimate $E1_{\text{PV}}$ of the above transition, various groups have used different many-body methods both in the linear response and sum-over-states approaches. By examining how these methods capture various electron correlation effects, we identify the underlying cause of sign discrepancies in the previously reported results. We also demonstrate how the double-core polarisation effects and scaled wave functions influence estimation of the $E1_{\text{PV}}$ amplitudes. The comprehensive discussions provided in this work will not only aid in our understanding on the potentials of the employed many-body methods but it will also serves as a road map for improving the $E1_{\text{PV}}$ calculation in the atomic systems further.
\end{abstract}

\date{}

\maketitle 

\section{Introduction}

There have been a long history of performing calculations of parity violating electric dipole ($E1_{\text{PV}}$) amplitude for the $6s ~ ^2S_{1/2} - 7s ~ ^2S_{1/2}$ transition in $^{133}$Cs by employing different state-of-the-art relativistic atomic many-body theories. In order to extract nuclear weak charge ($Q_W$) from atomic parity violation (APV) measurement, it is imperative to calculate $E1_{\text{PV}}$ very precisely (arguably less than 0.5\%). Among the early calculations, Dzuba et al had employed the time-dependent Hartree-Fock (TDHF) method \cite{Dzuba1984,Dzuba1985}, while M{\aa}rtensson had applied the combined coupled-perturbed Dirac-Hartree-Fock (CPDF) method and random-phase approximation (RPA), together referred as CPDF-RPA method \cite{maartensson1985}, to investigate the roles of core-polarization effects to $E1_{\text{PV}}$. Both these methods are technically equivalent, but M{\aa}rtensson had provided results at the intermediate levels using approximations at the Dirac-Hartree-Fock (DHF), CPDF and RPA methods as well as listing contributions from double-core-polarization (DCP) effects explicitly. Later, Blundell et al employed a linearized version relativistic coupled-cluster method in the singles and doubles excitation approximation (SD method) to estimate the $E1_{\text{PV}}$ amplitude of the above transition \cite{Blundell1992}. However, they adopted a sum-over-states approach in which matrix elements of the electric dipole (E1) operator and APV interaction Hamiltonian were evaluated for the transitions involving $np ~ ^2P_{1/2}$ intermediate states (called as ``Main" contribution) with the principal quantum number $n=6-9$. This method utilized the calculated E1 matrix elements and magnetic dipole hyperfine structure constants to estimate the uncertainty of $E1_{\text{PV}}$. Uncertainties from the energies were removed by considering the experimental energies, while contributions from core orbitals (referred to as ``Core" contribution henceforth) and higher $np ~ ^2P_{1/2}$ intermediate states (hereafter called as ``Tail" contribution) were estimated using lower-order methods. Following this work, Dzuba et al improved their calculation of TDHF method by incorporating correlation contributions through the Br\"uckner orbitals (BO) and referred the approach as RPA$+$BO method \cite{Dzuba2001}. Higher-order contributions from the Breit, lower-order QED and neutron skin effects were added subsequently through different works to claim for more precise $E1_{\text{PV}}$ value for extracting the beyond the Standard Model physics \cite{Dzuba1989, Milstein2002, Brown2009, Derevianko2001, Kozlov2001, Shabaev2005, Dzuba2002}.  Soon after these theoretical results, relativistic coupled-cluster (RCC) theory with singles and doubles approximation (RCCSD method) was employed to treat both the electromagnetic and weak interactions on an equal footing \cite{Sahoo2006, Sahoothesis, Sahoo2010}. 

A decade ago, Porsev et al made further improvement in the sum-over-states result of Blundell et al by considering contributions from the non-linear terms of the RCCSD method to their SD method as well as adding valence triple excitations (CCSDvT method) \cite{Porsev2010}. Their claimed accuracy to the $E1_{\text{PV}}$ amplitude of the $6s ~ ^2S_{1/2} - 7s ~ ^2S_{1/2}$ transition in $^{133}$Cs was about 0.27\%. However, the physical effects that were taken into account for the evaluation of the Core and Tail contributions were not explicitly stated and the Core and Tail contributions were still estimated using a combination of many-body methods. We collectively refer to these two contributions as X-factor and understanding accurate evaluation of this quantity is one of the objectives of the present work. To improve accuracy of the X-factor reported by Porsev et al., Dzuba et al. estimated this value in 2012 using their TDHF approach \cite{Dzuba2012} in the similar line to their earlier works \cite{Dzuba1984, Dzuba1985} but omitted the DCP contributions. However, this calculation pointed out about the opposite sign for the Core contribution in contrast to the earlier reported values. The opposite sign of the Core contribution of Dzuba et al with Porcev et al was criticized in two papers \cite{Safronova2018, Wieman2019}, which prompted for carrying out further investigation on different correlation contributions to $E1_{\text{PV}}$ from the first-principle approach. In 2021, Sahoo et al improved their calculation of the above $E1_{\text{PV}}$ amplitude by implementing the singles, doubles and triples approximation to both the unperturbed and perturbed wave functions (RCCSDT method) and using a much bigger set of basis functions \cite{Sahoo2021}. They also used the same $V^{N-1}$ potential as in the previous cases and presented the Core and Valence (Main and Tail together) contributions explicitly. As per the convention adopted in this approach, the Core contribution agreed with the earlier RCCSD result \cite{Sahoo2006, Sahoothesis, Sahoo2010}, and was close to the reported value of Blundell et al \cite{Blundell1992} and Porsev et al \cite{Porsev2010}. Subsequently, Roberts and Ginges argued again in favour of opposite sign of Core contribution than other findings by giving intermediate results of their RPA$+$BO method \cite{Roberts2022}. 

It is evident from the above discussions that the sign problem in the evaluation of the Core contribution to the $E1_{\text{PV}}$ amplitude of the $6s ~ ^2S_{1/2} - 7s ~ ^2S_{1/2}$ transition in $^{133}$Cs must be well understood. Also, Roberts in 2013 \cite{Roberts2013} reported different DCP contributions to above quantity by using the TDHF method than the CPDF-RPA values reported by Maartensson \cite{maartensson1985} {\it albeit} both the methods are equivalent. It also needs further verification. In this work, we attempt to discuss both the problems by demonstrating how various electron correlation affects are treated in all the previously employed methods to estimate $E1_{\text{PV}}$. For better comprehensive understanding, we also analyze the $E1_{\text{PV}}$ amplitude of the $6s ~ ^2S_{1/2} - 5d ~ ^2D_{3/2}$ transition in $^{133}$Cs. We first present various approaches from the general many-body theory point of view and relate them with the formulation adopted by the previously employed methods. It can help in finding how the definition of Core contribution changes depending on an approach. Further, we show how different correlation effects including the DCP of the TDHF and CPDF-RPA methods are subsumed within the RCCSD method approximation. 

\section{Theory \& Methodology}

We consider the Dirac-Coulomb (DC) Hamiltonian for the analysis, which in atomic units (a.u.) is given by
\begin{eqnarray}\label{eq:DC}
H &=&\sum_i \Lambda_i^{+}\left[c\mbox{\boldmath${\vec \alpha}$}^D \cdot {\vec p}_i+(\beta-1)c^2+V_{nuc}(r_i)\right]\Lambda_i^{+} +\sum_{i,j>i}\Lambda_i^{+}\Lambda_j^{+}\frac{1}{r_{ij}}\Lambda_i^{+}\Lambda_j^{+} \nonumber \\
&=& \sum_i h (r_i) + \sum_{i,j>i} g (r_{ij}),
\end{eqnarray}
where $c$ is the speed of light, $\mbox{\boldmath${\vec \alpha}$}^D$ and $\beta$ are the Dirac matrices, ${\vec p}$ is the single particle momentum operator, $\sum_{i,j}\frac{1}{r_{ij}}$ represents the two-body Coulomb potential between the electrons located at the $i^{th}$ and $j^{th}$ positions, and $\Lambda^{+}$ represents the projection operator which projects out the positive-energy orbitals when acts on the Dirac sea.

Next, we define the nuclear-spin-independent APV interaction Hamiltonian as
\begin{eqnarray} 
 H_{\rm PV}^{\rm NSI} &=& -\frac{G_F}{2\sqrt{2}} Q_W \gamma^5 \rho(r) , 
\end{eqnarray}
where $G_F= 2.22249 \times 10^{-14}$ a.u. is the Fermi constant, and $\rho(r)$ is the averaged nuclear density. In the presence of APV, the net atomic Hamiltonian is given by
\begin{eqnarray}
H_{at} &=& H + H_{\rm PV}^{\rm NSI} = H + G_F H_W ,
\end{eqnarray}
where $H_W = -\frac{ Q_W \gamma^5 \rho(r)}{2\sqrt{2}}$ is defined in order to treat $G_F$ as a small parameter to include contributions from $H_{\rm PV}^{\rm NSI}$ perturbatively through a many-body method. Using the wave functions of $H_{at}$, we determine $E1_{\text{PV}}$ of a transition between states $|\Psi_i \rangle $ and $|\Psi_f \rangle$ as
\begin{eqnarray}\label{defn}
E1_{\text{PV}} = \frac{\langle \Psi_f | D | \Psi_i \rangle} {\sqrt{\langle \Psi_f | \Psi_f \rangle \langle \Psi_i | \Psi_i \rangle } } , 
\label{eq1}
\end{eqnarray}
where $\Vec{D}=q\Vec{r}$ is the dipole or the E1 operator with the electron charge $q$ and position vector $\Vec r$. Since $G_F$ is very small, we can express atomic wave function of a state $|\Psi_v \rangle$ as
\begin{eqnarray}
|\Psi_v \rangle \simeq |\Psi_v^{(0)} \rangle + G_F |\Psi_v^{(1)} \rangle ,
\label{eq2}
\end{eqnarray}
where $|\Psi_v^{(0)} \rangle$ is the zeroth-order wave function containing contributions only from $H$ while $|\Psi_v^{(1)} \rangle$ includes one-order contribution from $H_W$ with respect to $|\Psi_v^{(0)} \rangle$. Substituting Eq. (\ref{eq2}) in Eq. (\ref{defn}) and keeping up to one-order terms in $G_F$, we get
\begin{eqnarray}
E1_{\text{PV}} & \simeq & G_F \left [ \frac{\langle \Psi_f^{(0)} | D | \Psi_i^{(1)} \rangle} {{\cal N}_{if} } +
\frac{\langle \Psi_f^{(1)} | D | \Psi_i^{(0)} \rangle} {{\cal N}_{if} } \right ] ,
\label{eq3}
\end{eqnarray}
where normalization factor ${\cal N}_{if}= \sqrt{N_f N_i}$ with $N_v=\langle \Psi_v^{(0)} | \Psi_v^{(0)} \rangle$. Results presented in this paper absorb $G_F$ value, so it will not appear explicitly here onwards. It can be noted that contribution from the first term is referred to as the initial perturbed state contribution whereas contribution from the second term is referred to as the final perturbed state contribution.

In the sum-over-states approach, the first-order wave function of a general state can be expressed as
\begin{eqnarray}
|\Psi_v^{(1)} \rangle &=&  \sum_{I \ne v} | \Psi_I^{(0)} \rangle \frac{ \langle \Psi_I^{(0)} | H_W | \Psi_v^{(0)} \rangle} {(E_v^{(0)} - E_I^{(0)}) } ,
\end{eqnarray}
where $| \Psi_I^{(0)} \rangle$ are the zeroth-order intermediate states and $E_n^{(0)}$ is the unperturbed energy of the $n^{th}$ level. Thus, Eq. (\ref{eq3}) can be written as
\begin{eqnarray}
E1_{\text{PV}} &=&  \sum_{I \ne i} \frac{\langle \Psi_f^{(0)} | D | \Psi_I^{(0)} \rangle \langle \Psi_I^{(0)} | H_W | \Psi_i^{(0)} \rangle} { (E_i^{(0)} - E_I^{(0)}) } \nonumber \\ && +
\sum_{I \ne f} \frac{\langle \Psi_f^{(0)} | H_W  | \Psi_I^{(0)} \rangle \langle \Psi_I^{(0)} | D| \Psi_i^{(0)} \rangle} {(E_f^{(0)} - E_I^{(0)}) }.
\label{eq4}
\end{eqnarray}

In the work by Porsev et al. \cite{Porsev2010}, the Core ($C$), Main ($V$) and Tail ($T$) contributions to the $E1_{\text{PV}}$ amplitude of the $6s ~ ^2S_{1/2} - 7s ~ ^2S_{1/2}$ transition in $^{133}$Cs were estimated as
\begin{eqnarray}
E1_{\text{PV}}(C) &=&  \sum_{n \le 5 } \frac{\langle 7S_{1/2} | D | nP_{1/2} \rangle \langle nP_{1/2} | H_W |6S_{1/2}  \rangle} {(E_{6S_{1/2}} - E_{nP_{1/2}} ) } \nonumber \\ & + &
\sum_{n \le 5} \frac{\langle 7S_{1/2} | H_W  | nP_{1/2} \rangle \langle nP_{1/2} | D| 6S_{1/2} \rangle} {(E_{7S_{1/2}} - E_{nP_{1/2}} )} \ \ \
\label{eq6}  \\
E1_{\text{PV}}(V) &=&  \sum_{n=6-9} \frac{\langle 7S_{1/2} | D | nP_{1/2} \rangle \langle nP_{1/2} | H_W |6S_{1/2}  \rangle} {(E_{6S_{1/2}} - E_{nP_{1/2}} ) } \nonumber \\ & + &
\sum_{n=6-9} \frac{\langle 7S_{1/2} | H_W  | nP_{1/2} \rangle \langle nP_{1/2} | D| 6S_{1/2} \rangle} {(E_{7S_{1/2}} - E_{nP_{1/2}} )} \ \ \
\label{eq7}  
\end{eqnarray}
and
\begin{eqnarray}
E1_{\text{PV}}(T) &=&  \sum_{n \ge 10 } \frac{\langle 7S_{1/2} | D | nP_{1/2} \rangle \langle nP_{1/2} | H_W |6S_{1/2}  \rangle} {(E_{6S_{1/2}} - E_{nP_{1/2}} ) } \nonumber \\ & + &
\sum_{n \ge 10} \frac{\langle 7S_{1/2} | H_W  | nP_{1/2} \rangle \langle nP_{1/2} | D| 6S_{1/2} \rangle} {(E_{7S_{1/2}} - E_{nP_{1/2}} )}, \ \ \
\label{eq8}    
\end{eqnarray}
respectively. Such classifications are based on the fact that $|\Psi_i^{(0)}\rangle$, $|\Psi_f^{(0)}\rangle$ and $|\Psi_I^{(0)}\rangle$ are represented by only single Slater determinants like $np ~ ^2P_{1/2}$ configurations. However, wave functions of multi-electron atomic systems are determined through a many-body method by expressing as a linear combination of many Slater determinants which can differ by either single or multiple entries of rows or columns. As a result, contributions from cross-terms involving other Slater determinants, e.g. excited configurations $5p^5 6s 7s$ of the intermediate states with respect to both the $6s ~ ^2S_{1/2}$ and $7s ~ ^2S_{1/2}$ states, cannot appear through the above breakup. These limitations can be overcome by obtaining solutions for the unperturbed and first-order perturbed wave functions as 
\begin{eqnarray}
H |\Psi_v^{(0)}\rangle = E_v^{(0)} |\Psi_v^{(0)}\rangle 
\label{eq0}
\end{eqnarray}
and 
\begin{equation}
(H-E_v^{(0)})|\Psi_v^{(1)}\rangle=(E_v^{(1)} -H_W)|\Psi_v^{(0)}\rangle ,
\label{eq1}
\end{equation}
respectively, where $E_v^{(1)}=0$ for odd-parity interaction operators.

Sometime $E1_{\text{PV}}$ is expressed in the following equivalent forms to evaluate it more conveniently 
\begin{eqnarray}
E1_{\text{PV}} &=& \frac{\langle \Psi_f^{(0)} | H_W | \tilde{\Psi}_i^{(1)} \rangle} {{\cal N}_{if} } + \frac{\langle \tilde{\Psi}_f^{(1)} | H_W | \Psi_i^{(0)} \rangle} {{\cal N}_{if} } , \label{eqn7} \\
  &\equiv& \frac{\langle \Psi_f^{(0)} | D | \Psi_i^{(1)} \rangle} {{\cal N}_{if} } +
\frac{\langle \Psi_f^{(0)} | H_W  | \tilde{\Psi}_i^{(1)} \rangle} {{\cal N}_{if} }  \label{eq8a} \\
 & \equiv& \frac{\langle \tilde{\Psi}_f^{(1)} | H_W | \Psi_i^{(0)} \rangle} {{\cal N}_{if} } +
\frac{\langle \Psi_f^{(1)} | D  | \Psi_i^{(0)} \rangle} {{\cal N}_{if} } ,
\label{eq8b}
\end{eqnarray}
by defining 
\begin{eqnarray}
|\tilde{\Psi}_i^{(1)} \rangle &=&  \sum_{I \ne f } | \Psi_I^{(0)} \rangle \frac{ \langle \Psi_I^{(0)} | D | \Psi_i^{(0)} \rangle} {(E_i^{(0)} - E_I^{(0)} + \omega ) }
\label{eq9}
\end{eqnarray}
and 
\begin{eqnarray}
|\tilde{\Psi}_f^{(1)} \rangle &=&  \sum_{I \ne i} | \Psi_I^{(0)} \rangle \frac{ \langle \Psi_I^{(0)} | D | \Psi_f^{(0)} \rangle} {(E_f^{(0)} - E_I^{(0)} - \omega ) } ,
\label{eq10}
\end{eqnarray}
where $\omega = E_f^{(0)} - E_i^{(0)}$ is the excitation energy between the initial and final states. The solutions for this modified first-order perturbed wave functions for the initial and final states are obtained by 
\begin{eqnarray}
(H-E_i^{(0)}-\omega) |\tilde{\Psi}_i^{(1)}\rangle & =&   - D |\Psi_i^{(0)}\rangle 
\label{eqd1}
\end{eqnarray}
and
\begin{eqnarray}
(H-E_f^{(0)}+\omega) |\tilde{\Psi}_f^{(1)}\rangle & =&  - D |\Psi_f^{(0)}\rangle .
\label{eqd2}
\end{eqnarray} 
One thing to note here is that if one uses experimental value of $\omega$ ($\omega^{ex}$) then it may introduce inequality in the energy denominator which will lead to inconsistency in the $E1_{\text{PV}}$ result. This can be shown clearly by the following equation
\begin{eqnarray}
E1_{\text{PV}} &=&  \sum_{I \ne i} \frac{\langle \Psi_f^{(0)} | D | \Psi_I^{(0)} \rangle \langle \Psi_I^{(0)} | H_W | \Psi_i^{(0)} \rangle} { (E_i^{(0)} - E_I^{(0)}-\delta\omega)} \nonumber \\ && +
\sum_{I \ne f} \frac{\langle \Psi_f^{(0)} | H_W  | \Psi_I^{(0)} \rangle \langle \Psi_I^{(0)} | D| \Psi_i^{(0)} \rangle} {(E_f^{(0)} - E_I^{(0)}+\delta\omega) }.
\end{eqnarray}
where, $\delta\omega=\omega^{ex}-\omega$ with $\omega$ being the theoretical value, cannot be zero when $\omega$ is obtained using a particular many-body method. 

Before applying particular methods of our choice to evaluate $E1_{\text{PV}}$, we demonstrate here general approaches using the Bloch's prescription to determine the unperturbed and first-order perturbed wave functions which would help us to connect contributions arising through different methods for comparative analyses.  We consider the DHF Hamiltonian $H_0$ to obtain mean-field wave function $| \Phi_0 \rangle$ of the closed core of Cs. Then, the DHF wave function of a state with valence orbital $v$ is defined as $| \Phi_v \rangle = a_v^{\dagger} | \Phi_0 \rangle$. To obtain the exact wave function of the state, we use the Bloch's prescription as \cite{Lindgren}
\begin{eqnarray}
 | \Psi_v \rangle = \Omega^v | \Phi_v \rangle ,
 \label{eqp}
\end{eqnarray} 
where $\Omega^v$ is known as the wave operator. In the $V^{N-1}$ potential approximation, we first solve electron correlation effects among electrons from the core orbitals of $| \Phi_0 \rangle$. Then, correlation effects involving electron from the valence orbital are included. Accordingly $\Omega^v$ is divided into two parts $\Omega^v = \Omega_0 + \Omega_v$, where $\Omega_0$ represents wave operator accounting correlations of electrons only from the core orbitals while $\Omega_v$ takes care of correlations of electrons from all orbitals including the valence orbital. In a given many-body method, we can solve amplitudes of the above wave operators using the equations
\begin{eqnarray}
 [\Omega_0, H_0] P_0 &=& U_{res} \Omega_0 P_0  
 \label{blw0}
\end{eqnarray}
and
\begin{eqnarray}
 [\Omega_v, H_0] P_v &=& U_{res} (\Omega_0 + \Omega_v) P_v   \nonumber \\
 && -  \Omega_v P_v U_{res} (\Omega_0 + \Omega_v) P_v,
\label{blwv}
\end{eqnarray}
where $P_n= |\Phi_n \rangle \langle \Phi_n |$ and $Q_n=1-P_n$ with $n\equiv 0,v$, and $U_{res}=H-H_0$ is known as the residual interaction that contributes to the amplitudes of $\Omega_0$ and $\Omega_v$. In fact, the energy of the $|\Psi_v \rangle$ state can be evaluated as the expectation value of the effective Hamiltonian 
\begin{eqnarray}
H_{eff} = P_v U_{res} (\Omega_0 + \Omega_v) P_v 
\end{eqnarray}
with respect to the reference state $|\Phi_v \rangle$. It should be noted that energy of the state is given by
\begin{eqnarray}
    E_v=\langle\Phi_v|H_{eff}|\Phi_v\rangle .
\end{eqnarray}

For the choice of $V^N$ potential in the generation of single particle orbitals, amplitudes of the $\Omega^v$ operator can be estimated using 
$[\Omega^v, H_0] P_v = U_{res}  \Omega^v P_v$.
It means that $E_v$ does not appear in the wave function determining equation for the case of $V^N$ potential. Thus, the core-valence interaction effects in the construction of DHF potential in case of $V^{N-1}$ is partly compensated through the wave operator amplitude determining equation through the extra term with energy. If any method utilizes the $V^{N-1}$ potential without taking into account the above-mentioned extra term, it should be termed as an improper formulation of the many-body theory as have been adopted in some of the previous works discussed in the first section.

Using the wave operator formalism, we can obtain solutions of Eqs. (\ref{eq0}), 
 (\ref{eq1}), (\ref{eqd1}) and (\ref{eqd2}) as
\begin{eqnarray}
 | \Psi_v^{(0)} \rangle &=&\Omega^{v(0)}| \Phi_v \rangle  = (\Omega_0^{(0)} + \Omega_v^{(0)} ) | \Phi_v \rangle  , \nonumber \\
 | \Psi_v^{(1)} \rangle &=&\Omega^{v(1)}| \Phi_v \rangle = (\Omega_0^{(1)} + \Omega_v^{(1)} ) | \Phi_v \rangle .
 \label{eq03}
\end{eqnarray}
and
\begin{eqnarray}
| \tilde{\Psi}_v^{(1)} \rangle &=& (\tilde{\Omega}_0^{(1)} + \tilde{\Omega}_v^{(1)} ) | \Phi_v \rangle  .
\end{eqnarray}

We can define the Core contribution in the $V^{N-1}$ potential approximation as 
\begin{eqnarray}
E1_{\text{PV}} (C) = \frac{\langle \Phi_0 | a_f [\Omega_0^{(0)\dagger} D \Omega_0^{(1)} + \Omega_0^{(1)\dagger} D \Omega_0^{(0)} ] a_i^{\dagger} | \Phi_0 \rangle} {{\cal N}_{if}} . \ \ \
\label{eqcc}
\end{eqnarray}
Thus, the Valence contribution can be obtained from the rest of the terms as 
\begin{eqnarray}
E1_{\text{PV}} (V+T) = \frac{\langle \Phi_0 | a_f [\Omega_f^{(0)\dagger} D \Omega_0^{(1)} + \Omega_f^{(1)\dagger} D \Omega_0^{(0)} ] a_i^{\dagger} | \Phi_0 \rangle} {{\cal N}_{if}} \nonumber \\
+\frac{\langle \Phi_0 | a_f [\Omega_0^{(0)\dagger} D \Omega_i^{(1)} + \Omega_0^{(1)\dagger} D \Omega_i^{(0)} ] a_i^{\dagger} | \Phi_0 \rangle} {{\cal N}_{if}} \nonumber \\
+ \frac{\langle \Phi_0 | a_f [\Omega_f^{(0)\dagger} D \Omega_i^{(1)} + \Omega_f^{(1)\dagger} D \Omega_i^{(0)} ] a_i^{\dagger} | \Phi_0 \rangle} {{\cal N}_{if}} . \ \ \ \
\label{eqvc}
\end{eqnarray}
It follows that there cannot be any ambiguity in defining Core and Valence contributions with the help of wave operators. It also implies that the Core and Valence contributions for expressions given by Eqs. (\ref{eqn7}), (\ref{eq8a}) and (\ref{eq8b}) can be defined accordingly by replacing the respective wave operators with their counter tilde forms. To demonstrate how a Core correlation contribution from Eq. (\ref{eq4}) switches to Valence correlation contribution in their equivalent forms, we express the first-order perturbed wave operators in terms of unperturbed wave operators of the intermediate states. Thus, Eq. (\ref{eq4}) can be expressed as
\begin{eqnarray}
E1_{\text{PV}} &=&  \sum_{I \ne i} \frac{\langle \Phi_0 | a_f (\Omega_0^{(0)} + \Omega_f^{(0)})^{\dagger} D (\Omega_0^{(0)} + \Omega_I^{(0)}) a_I^{\dagger}| \Phi_0 \rangle} {{\cal N}_{if}} \nonumber \\
&\times& \frac{\langle \Phi_0 | a_I (\Omega_0^{(0)} + \Omega_I^{(0)})^{\dagger} H_W (\Omega_0^{(0)} + \Omega_i^{(0)}) a_i^{\dagger}| \Phi_0 \rangle} {(E_i^{(0)} - E_I^{(0)})} \nonumber \\
&+& \sum_{I \ne f} \frac{\langle \Phi_0 | a_f (\Omega_0^{(0)} + \Omega_f^{(0)})^{\dagger} H_W (\Omega_0^{(0)} + \Omega_I^{(0)}) a_I^{\dagger}| \Phi_0 \rangle} {{\cal N}_{if}} \nonumber \\
&\times& \frac{\langle \Phi_0 | a_I (\Omega_0^{(0)} + \Omega_I^{(0)})^{\dagger} D (\Omega_0^{(0)} + \Omega_i^{(0)}) a_i^{\dagger}| \Phi_0 \rangle} {(E_f^{(0)} - E_I^{(0)})} .
\label{eq50}
\end{eqnarray}
One of the Core contributing terms from this expression can be given by
\begin{eqnarray}
E1_{\text{PV}}^a &=& \sum_{I \ne i} \frac{\langle \Phi_0 | a_f \Omega_0^{(0) \dagger} D  \Omega_I^{(0)} a_I^{\dagger} a_I \Omega_I^{(0) \dagger} H_W \Omega_0^{(0)} a_i^{\dagger}| \Phi_0 \rangle} {(E_i^{(0)} - E_I^{(0)})} . \ \ \ \
\end{eqnarray}
Though $I \ne i$ in the above term, but the intermediate states can include valence orbital $f$ of the system such as $5p^5 6s 7s$ in Cs. Note that $\Omega$ operators contain, singly doubly, triply etc. excitation configurations. As a result, some contributing configurations through the above expression will be given by  
\begin{eqnarray}
E1_{\text{PV}}^b &=& \sum_{I \ne i} \frac{\langle \Phi_0 | a_f \Omega_f^{(0) \dagger} D  \Omega_0^{(0)} a_I^{\dagger} a_I\Omega_0^{(0) \dagger} H_W \Omega_0^{(0)} a_i^{\dagger}| \Phi_0 \rangle} {(E_f^{(0)} - E_I^{(0)} - \omega)} . \ \ \ \
\end{eqnarray}
It should be noted that the number of particles involved are same in the above expressions and $E1_{\text{PV}}^b$ can correspond to Valence contribution of $E1_{\text{PV}}$ through the expressions given by Eqs. (\ref{eqn7}), (\ref{eq8a}) and (\ref{eq8b}). We shall demonstrate explicitly next using examples of specific many-body methods of our interest to show how definitions of such correlation processes can interchange depending on whether $H_W$ or $D$ is taken as perturbation. However, the final result from a given method should be independent of which of them is treated as perturbation.

To show this point explicitly, we present results from the relativistic many-body perturbation theory (RMBPT) at the third-order (RMBPT(3) method) approximation. Assuming external perturbative operator in this method as $H_W$ (denoted by RMBPT$^w$ method) and $D$ (denoted by RMBPT$^d$ method) independently, we show how a few Core contributing terms of the RMBPT$^w$ method correspond to Valence contributing terms (and vice versa) in the RMBPT$^d$ method. 

We start our calculation by obtaining single particle orbitals using the modified DHF Hamiltonian and residual interaction as $F = \sum_i f (r_i) = \sum_i (h_i +  u_0 (r_i))$ and $V_{res}=H-F$ respectively in the DC approximation. Hence, the DHF orbital wave function for the valence orbital is obtained by 
\begin{eqnarray}
  f | v \rangle = \epsilon_v | v \rangle
\label{eqhfe}
\end{eqnarray}
with the single particle energies $\epsilon_i$ giving unperturbed DHF energy ${\cal E}_0 = \sum_b^{N_c} \epsilon_b $ and the DHF potential $U_{DHF} = \sum_i u_0 (r_i) $ is given by 
\begin{eqnarray}
u_0 | v \rangle = \sum_b^{N_c} \left [ \langle b | g  |b \rangle |v \rangle -  \langle b | g  |v \rangle |b \rangle \right ] 
\label{eqhfu}
\end{eqnarray}
for $b$ summing over all occupied-orbitals $N_c$. In the DHF method, both Eqs. (\ref{eqhfe}) and (\ref{eqhfu}) are solved iteratively to obtain self-consistent solutions. It can be followed from the above expression that for the determinant expressed by $|\Phi_k \rangle = a_k^{\dagger} |\Phi_0 \rangle$, the DHF energy is given by ${\cal E}_k = {\cal E}_0 + \epsilon_k$. Using wave functions from the DHF method, we can evaluate $E1_{\text{PV}}$ in the mean-field approach as 
\begin{eqnarray}
E1_{\text{PV}} = \langle \Phi_f | D | \Phi_i^{(1)} \rangle + \langle \Phi_f^{(1)} | D | \Phi_i \rangle,
\end{eqnarray}
where $| \Phi_{n=i,f}^{(1)} \rangle $ are the first-order perturbed wave functions with respect to $| \Phi_{n=i,f} \rangle $. 

\begin{figure}[t!]
\centering
\includegraphics[height=40mm, width=80mm]{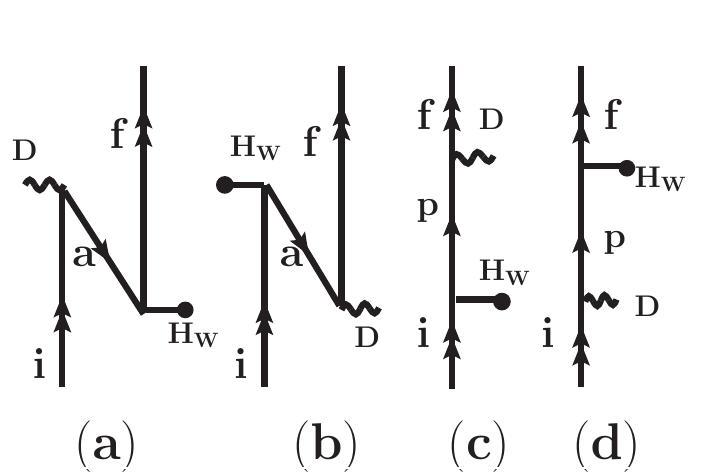}
\caption{Goldstone diagrams representing Core (a and b) and Valence (c and d) contributions to $E1_{\text{PV}}$ in the DHF method. Here, double arrows represent initial ($i$) and final ($f$) valence orbitals, single arrows going down ($a$) means occupied orbitals and single arrows going up ($p$) means virtual orbitals. The operators $D$ and $H_W$ are shown in curly line and a line with bullet point respectively.}
    \label{DF}
\end{figure}

Expanding $| \Phi_{n=i,f}^{(1)} \rangle $ in terms of intermediate states $|\Phi_I \rangle$, it yields 
\begin{eqnarray}
E1_{\text{PV}} &=& \sum_{I\ne i} \frac{\langle \Phi_f | D | \Phi_I \rangle \langle \Phi_I | H_W | \Phi_i \rangle} { {\cal E}_i - {\cal E}_I } \nonumber \\
  && + \sum_{I\ne f} \frac{\langle \Phi_f | H_W | \Phi_I \rangle \langle \Phi_I | D | \Phi_i \rangle} { {\cal E}_f - {\cal E}_I }  \nonumber \\
  & \equiv & \langle \Phi_f | H_W | \tilde{\Phi}_i^{(1)} \rangle + \langle \tilde{\Phi}_f^{(1)} | H_W | \Phi_i \rangle  \nonumber \\
    & \equiv &  \langle \Phi_f | H_W | \tilde{\Phi}_i^{(1)} \rangle + \langle \Phi_f | D | \Phi_i^{(1)} \rangle  \nonumber \\
     & \equiv & \langle \Phi_f^{(1)} | D | \Phi_i \rangle + \langle \tilde{\Phi}_f^{(1)} | H_W | \Phi_i \rangle ,
\end{eqnarray}
where tilde symbols are used for the perturbed wave function due to $D$. Using $H_W = \sum_i h_w (r_i)$ and $D=\sum_i d(r_i)$, the above expressions can be given in terms of single particle matrix elements as
\begin{eqnarray}
 E1_{\text{PV}} &=&  \sum_{a } \frac{ \langle f | d | a \rangle \langle a | h_w | i \rangle} {\epsilon_i - \epsilon_a } 
   + \sum_{a } \frac{\langle f | h_w | a \rangle \langle a | d | i \rangle} {\epsilon_f - \epsilon_a }  \nonumber \\
   &+&  \sum_{p \ne i} \frac{ \langle f | d | p \rangle \langle p | h_w | i \rangle} {\epsilon_i - \epsilon_p } 
   + \sum_{p \ne f} \frac{\langle f | h_w | p \rangle \langle p | d | i \rangle} {\epsilon_f - \epsilon_p } , \ \ \ \ 
\end{eqnarray}
where $|k=a,p \rangle$ denotes $k^{th}$ single particle DHF orbital with energy $\epsilon_k$ for the occupied orbitals denoted by $a$ and virtual orbitals denoted by $p$. In terms of wave operators, the DHF expression can be given by
\begin{eqnarray}
E1_{\text{PV}} &=& \langle \Phi_f | D \Omega^{i(0,1)} | \Phi_i \rangle + \langle \Phi_f | \Omega^{f(0,1) \dagger} D | \Phi_i \rangle \nonumber \\
      &\equiv & \langle \Phi_f |  \tilde{\Omega}^{f(0,1) \dagger} H_W | \Phi_i \rangle + \langle \Phi_f | H_W \tilde{\Omega}^{i(0,1)} | \Phi_i \rangle \nonumber \\
      & \equiv & \langle \Phi_f | D \Omega^{i(0,1)} | \Phi_i \rangle + \langle \Phi_f | H_W \tilde{\Omega}^{i(0,1)} | \Phi_i \rangle  \nonumber \\
     & \equiv & \langle \Phi_f |  \tilde{\Omega}^{f(0,1) \dagger} H_W | \Phi_i \rangle  + \langle \Phi_f | \Omega^{f(0,1) \dagger} D | \Phi_i \rangle , \ \ \ \
    \label{wvdf}
\end{eqnarray}
where $\Omega^{v(0,1)} = \Omega_0^{(0,1)} + \Omega_v^{(0,1)}$. For single excitations, $\Omega_{0}^{(0,1)} \rightarrow \Omega_{1}^{(0,1)}= \sum_{a,p} \frac{ \langle \Phi_a^p | H_W | \Phi_0 \rangle} { {\cal E}_0 - {\cal E}_a^p } a_p^{\dagger} a_a = \sum_{a,p} \frac{ \langle p | h_w | a \rangle} { \epsilon_a - \epsilon_p} a_p^{\dagger} a_a  \equiv \sum_{a,p} \Omega_a^p $ and $\Omega_{v}^{(0,1)} \rightarrow \Omega_{1v}^{(0,1)}= \sum_{p} \frac{ \langle \Phi_v^p | H_W | \Phi_v \rangle} { {\cal E}_v-{\cal E}_v^p} a_p^{\dagger} a_v=  \frac{ \langle p | h_w | v \rangle} { \epsilon_v - \epsilon_p} a_p^{\dagger} a_v \equiv  \sum_p \Omega_v^p $ for the intermediate states $|\Phi_a^p\rangle$ with energies ${\cal E}_a^p  = {\cal E}_0 + \epsilon_p - \epsilon_a$ and $|\Phi_v^p\rangle$ with energies ${\cal E}_v^p= {\cal E}_0 + \epsilon_p $ with respect to the $|\Phi_0\rangle$ and $|\Phi_v \rangle$ respectively. It should be noted that for double excitations $\Omega_{0}^{(0,1)} \rightarrow \Omega_{2}^{(0,1)}=0$ and $\Omega_{v}^{(0,1)} \rightarrow \Omega_{2v}^{(0,1)}=0$.
Figs. \ref{DF}(a) and (b) are the Goldstone diagrams corresponding to lowest-order Core contributing terms, while Figs. \ref{DF}(c) and (d) are the Goldstone diagrams representing the lowest-order Valence contributing terms. The Core and Valence contributions at the lowest-order are same whether wave functions are perturbed due to $H_W$ and $D$, because both $H_W$ and $D$ operators are the one-body operators, and only the singly excited intermediate states contribute to the DHF method. However, they can be different if one includes correlation effects through higher-level excitation configurations. This is because in an equivalent form $ D \Omega^{i(1)} \rightarrow \tilde{\Omega}^{f(1) \dagger} H_W$, so when doubly excited configurations of initial state can appear as singly excited configuration of the final state through the wave operators. This statement can be asserted with the help of the RMBPT(3)$^w$ and RMBPT(3)$^d$ methods at the lower-order level, and using the CPDF and RPA methods at the all-order level.   

\begin{figure}[t]
    \centering
\includegraphics[height=30mm,width=85mm]{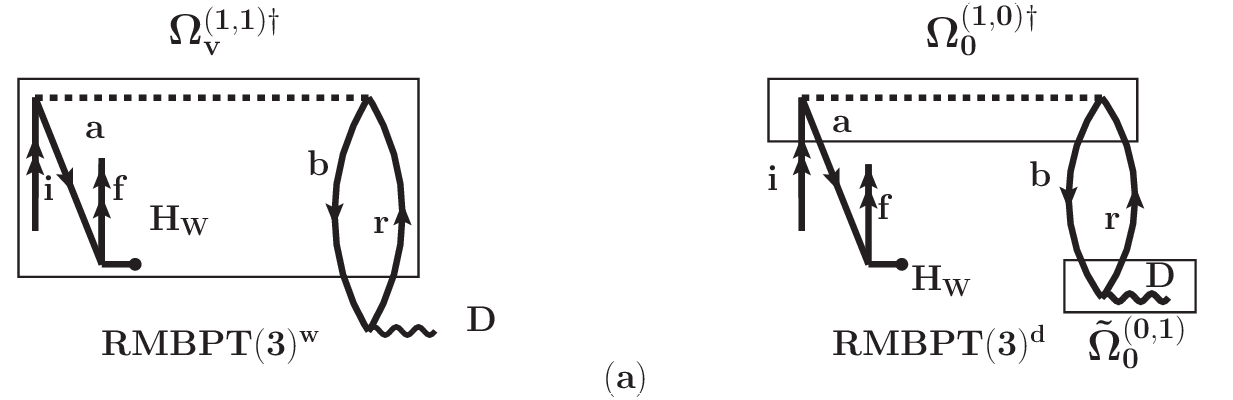}\\
\vskip0.3cm 
\includegraphics[height=30mm,width=85mm]{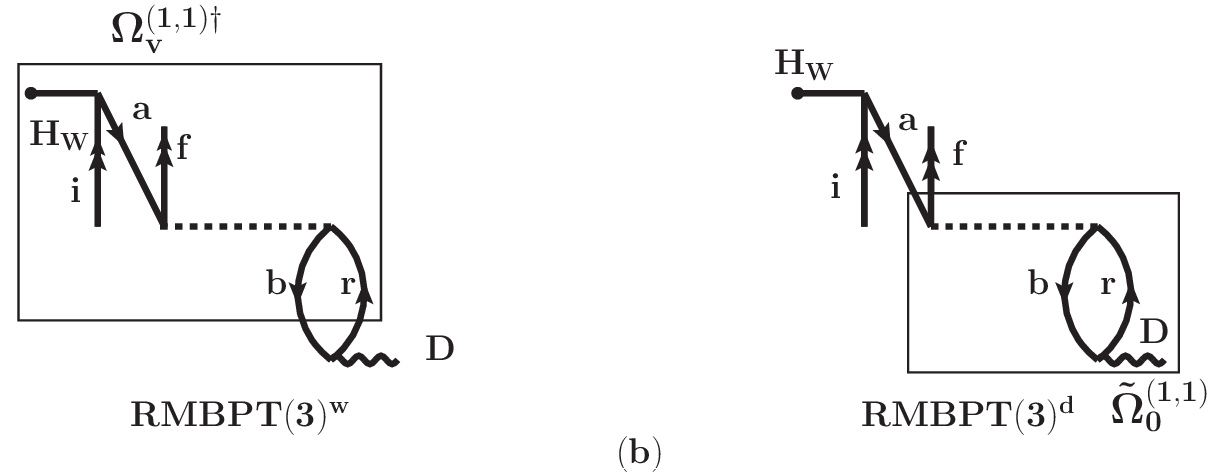}
\caption{A few typical Goldstone diagrams of the RMBPT(3) method. It demonstrates how Valence contributions through $\Omega_v^{(1,1)\dagger}D$ in the RMBPT(3)$^w$ method correspond to Core contributions through (a) $\Omega_0^{(1,0)\dagger} H_W \tilde{\Omega}_0^{(0,1)}$ and (b) $H_W \tilde{\Omega}_0^{(1,1)} $ in the RMBPT(3)$^d$ method.}
    \label{MBPT}
\end{figure}

In the RMBPT(3)$^w$ method, we express as
\begin{equation}
    H_{at} = F + \lambda_1 V_{res} + \lambda_2 H_W ,
\end{equation}
where $\lambda_1$ and $\lambda_2$ are arbitrary parameters introduced to count orders of $V_{res}$ and $H_W$ in the calculation. Hence, the unperturbed wave operators can be given by $\Omega_n^{(0)} = \sum_k \Omega_n^{(k,0)} $. Similarly, the first-order perturbed wave operators in the RMBPT(3)$^w$ method can be denoted by 
$\Omega_n^{(1)} = \sum_k \Omega_n^{(k,1)}$,
where subscript $n=0,v$ stands for core or valence operators and superscript $k$ denotes for order of $V_{res}$. Amplitudes of these wave operators using the Bloch's equation with the lower-order unperturbed wave operator $\Omega_n^{(0,0)}=1$. In the RMBPT(3)$^d$ method, $\tilde{\Omega}_0^{(1)}$, $\tilde{\Omega}_i^{(1)}$ and $\tilde{\Omega}_f^{(1)}$ operators have to be used by replacing $H_W$ by $D$ in the above equations. We consider a few Goldstone diagrams in Fig. \ref{MBPT} that represent Valence (Main and Tail) contributions in the RMBPT(3)$^w$ method to demonstrate how they turn to Core contributing diagrams in the RMBPT(3)$^d$ method.
\begin{figure}[t]
    \centering
    \includegraphics[height=40mm,width=85mm]{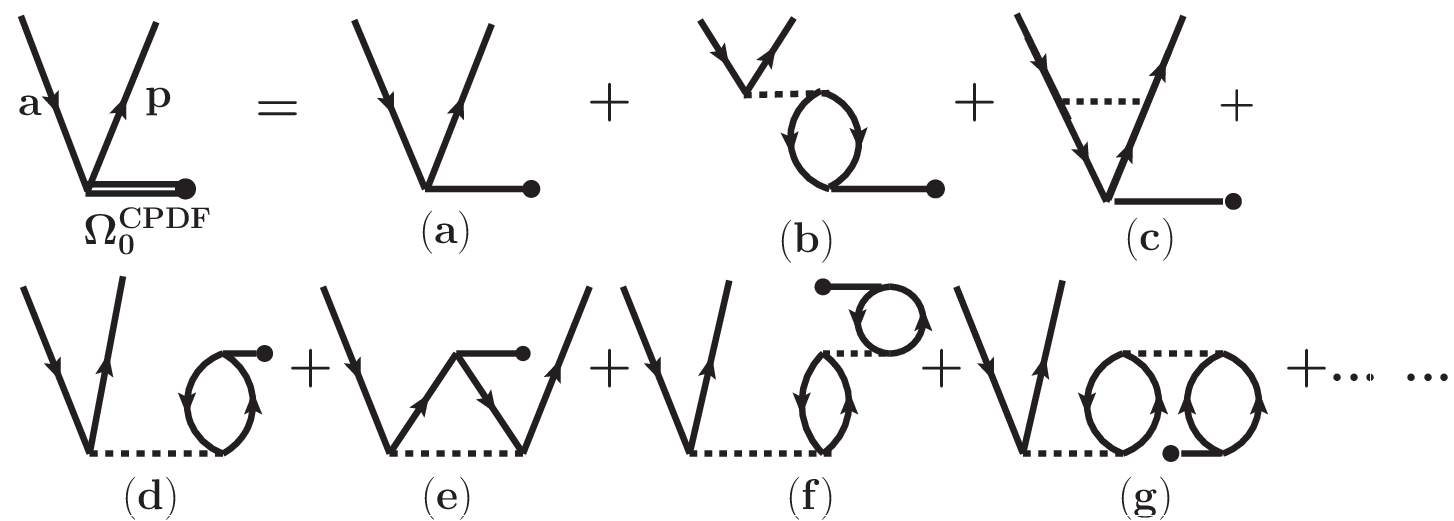}
    \caption{Goldstone diagrams contributing to amplitude determining equation of $\Omega_0^{CPDF}$. Through iterative scheme these effects are included to all-orders in the CPDF method.}
    \label{CPHF1}
\end{figure}

Next, we extend the above discussion considering the CPDF and RPA methods. In the CPDF method, the unperturbed wave function is taken as the DHF orbital wave function of the valence electrons while the perturbed wave functions are obtained as
\begin{eqnarray}
(f_v - \epsilon_v) |v^{PV} \rangle =  -h_w |v \rangle - u_v^{PV}  |v \rangle , 
\label{eqhf1}
\end{eqnarray}
where 
\begin{eqnarray}
  u_v^{PV} | v \rangle = \sum_b^{N_c} \left [ \langle b | g  |b^{PV} \rangle |v \rangle -  \langle b | g  |v \rangle |b^{PV}\rangle \right. \nonumber  \\ 
  \left. + \langle b^{PV} | g  |b \rangle |v \rangle -  \langle b^{PV} | g  |v \rangle |b \rangle \right ] .
\label{eqhfu1}
\end{eqnarray}

\begin{figure}[t]
    \centering
    \includegraphics[height=4.0cm,width=8.0cm]{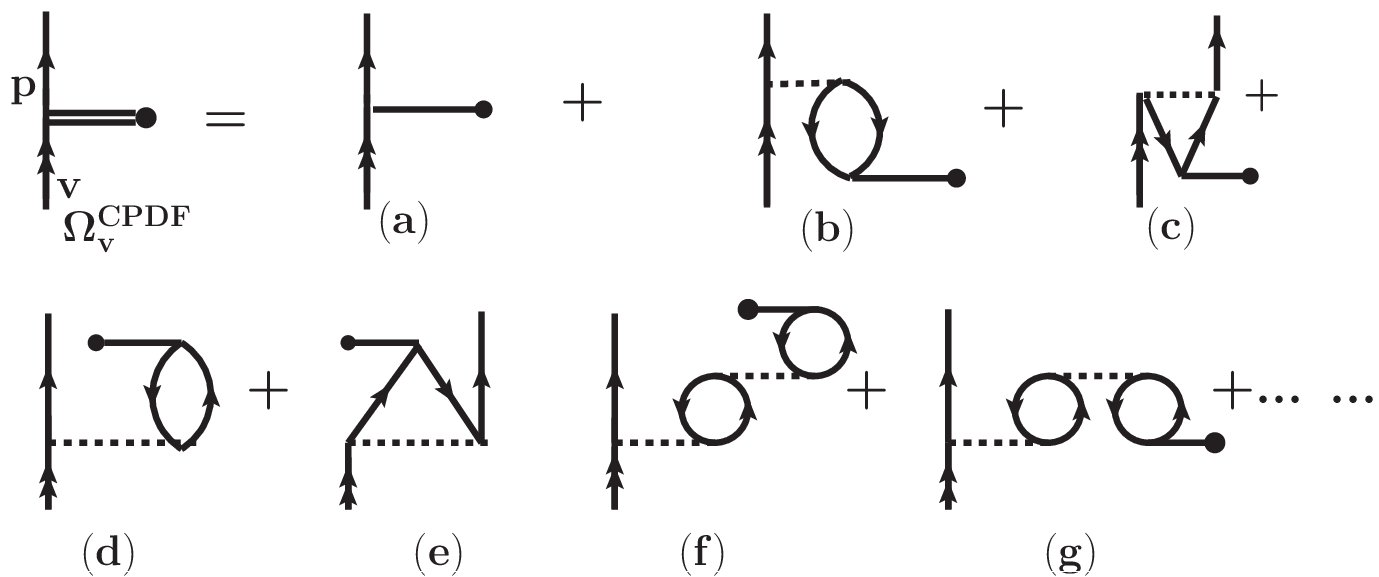}
    \caption{Diagrams denoting amplitude solving equation for $\Omega_v^{CPDF}$. These core-polarization effects are included to all-orders in the CPDF method.}
    \label{CPHF2}
\end{figure}

Both Eqs. (\ref{eqhf1}) and (\ref{eqhfu1}) are solved iteratively to obtain the self-consistent solutions to account for core-polarization effects to all-orders. Using the above APV modified orbitals, $E1_{\text{PV}}$ can be evaluated as  
\begin{eqnarray}
E1_{\text{PV}} &=& \langle f|d|i^{PV} \rangle + \langle f^{PV} |d|i\rangle  \nonumber \\
    &=& \langle \Phi_f | D \Omega^{i,CPDF} | \Phi_i \rangle + \langle \Phi_f | \Omega^{f,CPDF \dagger} D | \Phi_i \rangle, \ \ \
\label{eqcphf}
\end{eqnarray}
where $\Omega^{v,CPDF} = \Omega_0^{CPDF} + \Omega_{v}^{CPDF} = \sum_{k=1}^{\infty} \left [\sum_{a,p} \Omega_{a,p}^{(k,1)} + \sum_p \Omega_{v,p}^{(k,1)} \right ] $. The Goldstone diagrams contributing to the amplitudes of $\Omega_0^{CPDF}$ are shown in Fig. \ref{CPHF1}. Similarly, the Goldstone diagrams contributing to the amplitudes of $\Omega_v^{CPDF}$ are shown in Fig. \ref{CPHF2}. It is easy to follow from these diagrams that how the core-polarization effects are included to all-orders through the CPDF method but they completely miss out pair-correlation contributions. 

\begin{figure}[t]
    \centering
    \includegraphics[height=40mm,width=87mm]{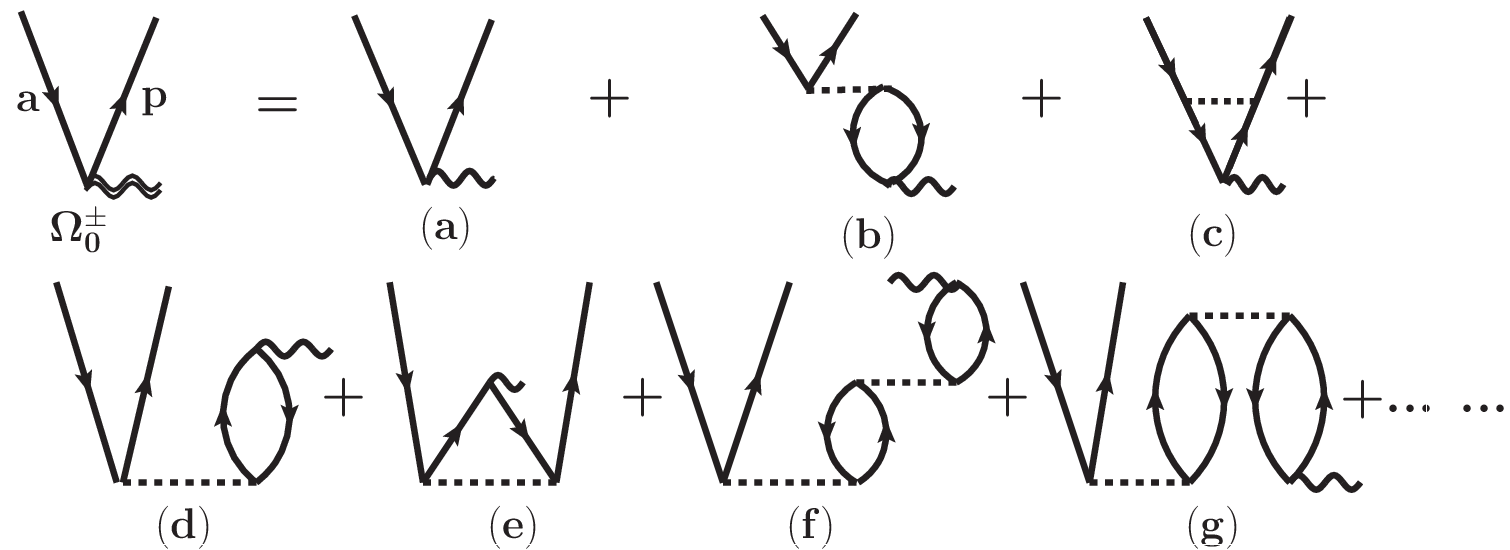}
    \caption{Graphical representation of $\Omega_0^{\pm}$ and its expansion in terms of lower-order RMBPT(3) method.}
    \label{RPA1}
\end{figure}

\begin{figure}[t]
    \centering
    \includegraphics[height=40mm,width=85mm]{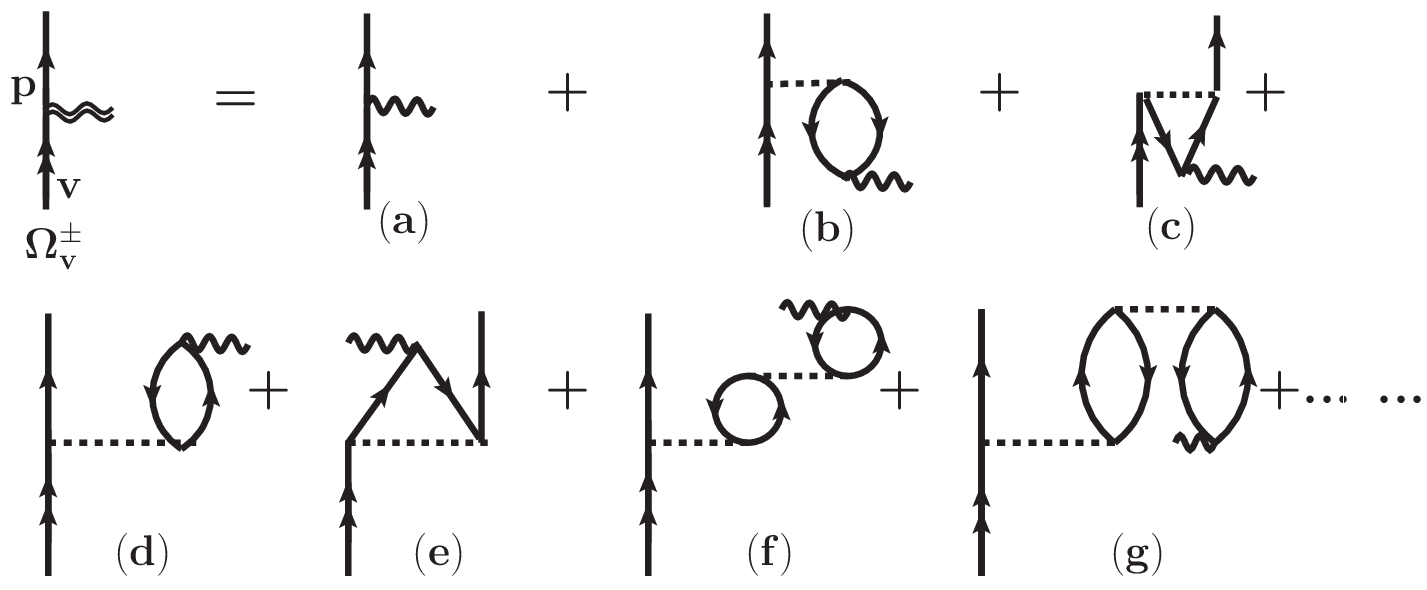}
 \caption{Graphical representation of $\Omega_v^{\pm}$ and its expansion in terms of lower-order RMBPT(3) method.}
    \label{RPA2}
\end{figure}

The CPDF method does not include correlation effects through the $D$ operator. It is possible to capture core-polarization effects through the $D$ operator in the RPA method by proceeding in the similar manner to obtain the perturbed valence orbital wave functions as
\begin{eqnarray}
(f_v - \epsilon_v \mp \omega) |v^{\pm} \rangle =  - d |v \rangle - u_v^{\pm}  |v \rangle , 
\label{eqhf2}
\end{eqnarray}
where 
\begin{eqnarray}
  u_v^{\pm} | v \rangle = \sum_b^{N_c} \left [ \langle b | g  |b^{\pm}\rangle |v \rangle -  \langle b | g  |v \rangle |b^{\pm} \rangle \right. \nonumber  \\ 
  \left. + \langle b^{\mp} | g |b \rangle |v \rangle - \langle b^{\mp} | g |v \rangle |b \rangle \right ] .
\label{eqhfu2}
\end{eqnarray}
It yields
\begin{eqnarray}
E1_{\text{PV}} &=& \langle f^-|h_w|i \rangle + \langle f |h_w|i^+\rangle  \nonumber \\
    &=& \langle \Phi_f | \Omega^{f,- \dagger} H_W | \Phi_i \rangle + \langle \Phi_f | H_W \Omega^{i,+} | \Phi_i \rangle , \ \ \
\label{eqcphf}
\end{eqnarray}
where $\Omega^{v,\pm} = \Omega_0^{\pm} + \Omega_v^{\pm} = \sum_{k=1}^{\infty} \left [\sum_{a,p} \Omega_{a,p}^{\pm(k,1)} + \sum_p \Omega_{v,p}^{\pm(k,1)} \right ] $. The Goldstone diagrams contributing to the amplitude determining equation for Core operator are shown in Fig. \ref{RPA1}, while the diagrams contributing to the amplitudes of the Valence operator are shown in Fig. \ref{RPA2}.

It can be followed that the Core correlations (without DHF contributions) arising in the RPA are distinctly different than that appear via the CPDF method. It shows that the Core contributing diagrams of the CPDF method correspond to the lower-order Core contributing diagrams of the RMBPT(3)$^w$ method. Similarly,  the Core contributing diagrams of RPA correspond to the lower-order Core contributing diagrams of the RMBPT(3)$^d$ method. Now relating these diagrams to the RMBPT(3) diagrams shown in Fig. \ref{MBPT}, it can be easily understood that some of the Core contributions arising through the CPDF method and RPA in the combined CPDF-RPA method cannot correspond to the Core contributions when either $H_W$ or $D$ is treated as perturbation for the evaluation of $E1_{\text{PV}}$ in the RCCSD method. This point can be shown categorically using the CPDF-RPA and RCCSD methods as described below. 

To include core-polarization effects through both the bra and ket states, it is necessary to include the $H_W$ and $D$ operators in the perturbation. For this purpose, we define the total Hamiltonian as
\begin{eqnarray}
H_t &=& H + \lambda_2 H_W + \lambda_3 D
 \equiv H_{int} + \lambda_3 D .
\end{eqnarray}
The exact atomic wave function ($| \overline{\Psi}_v \rangle$) of $H_t$ can be expressed as
\begin{eqnarray}
| \overline{\Psi}_v \rangle &=& |\Psi_v^{(0,0)} \rangle + \lambda_2 |\Psi_v^{(1,0)} \rangle + \lambda_3 | \tilde{\Psi}_v^{(0,1)} \rangle \nonumber \\
  && + \lambda_2 \lambda_3 |\Psi_v^{(1,1)} \rangle + \cdots ,
\end{eqnarray}
where $|\Psi_v^{(m,n)} \rangle $ considers $m$ orders of $H_W$ and $n$ orders of $D$. In this case, we can determine the $E1_{\text{PV}}$ amplitude as the transition amplitude of $O\equiv \lambda_2 H_W + \lambda_3 D $ between the initial perturbed state to the final unperturbed state or between the initial unperturbed state to the final perturbed state (see Eqs. (\ref{eq8a}) and (\ref{eq8b})). i.e. 
\begin{eqnarray}
E1_{\text{PV}} &=&  \langle \Psi_f^{(0,0)} |\Psi_i^{(1,1)} \rangle + \langle \Psi_f^{(0,0)} |  D |\Psi_i^{(1,0)} \rangle \nonumber \\ && + \langle \Psi_f^{(0,0)} |  H_W | \tilde{\Psi}_i^{(0,1)} \rangle \nonumber \\
   &=& \langle \Phi_f | \Omega_f^{(0,0)\dagger} \Omega_i^{(1,1)} | \Phi_i \rangle + \langle \Phi_f | \Omega_f^{(0,0)\dagger}  D \Omega_i^{(1,0)}  |\Phi_i \rangle \nonumber \\ && + \langle \Phi_f | \Omega_f^{(0,0)\dagger} H_W \tilde{\Omega}_i^{(0,1)} |\Phi_i \rangle 
\label{cprpa1}
 \end{eqnarray}
 or
  \begin{eqnarray}
  E1_{\text{PV}} &=&  \langle \Psi_f^{(1,1)} |\Psi_i^{(0,0)} \rangle + \langle \Psi_f^{(1,0)} |  D |\Psi_i^{(0,0)} \rangle \nonumber \\ && + \langle \tilde{\Psi}_f^{(0,1)} |  H_W |\Psi_i^{(0,0)} \rangle \nonumber \\
   &=& \langle \Phi_f | \Omega_f^{(1,1)\dagger} \Omega_i^{(0,0)} | \Phi_i \rangle + \langle \Phi_f | \Omega_f^{(1,0)\dagger}  D \Omega_i^{(0,0)}  |\Phi_i \rangle \nonumber \\ && + \langle \Phi_f | \tilde{\Omega}_f^{(0,1)\dagger} H_W \Omega_i^{(0,0)} |\Phi_i \rangle , 
\label{cprpa2}
 \end{eqnarray}
keeping terms that are of the order of $\lambda_2 \lambda_3$. It is important to use both expressions in an approximated method to verify numerical uncertainty. If $\omega^{ex}$ is used, the results from both equations may not agree due to inconsistencies in the treatment of the intermediate states as was shown earlier.

The modified single particle Hamiltonian for the corresponding Hamiltonian $H_t$ in the CPDF-RPA method are obtained by
\begin{eqnarray}
(f_v - \epsilon_v \mp \omega) |v^{PV\pm} \rangle &=&  -d |v^{PV} \rangle - u_v^{\pm} | v^{PV} \rangle - h_w |v^{\pm} \rangle \nonumber \\ && - u_v^{PV} | v^{\pm} \rangle - u_v^{PV \pm}  |v \rangle , 
\label{eqhf4}
\end{eqnarray}
where 
\begin{eqnarray}
  u_v^{PV \pm} | v \rangle = \sum_b^{N_c} \left [ \langle b^{\mp} | g  |b^{PV} \rangle |v \rangle   -  \langle b^{\mp} | g  |v \rangle |b^{PV} \rangle \right. \nonumber  \\ 
  \left. + \langle b^{PV} | g  |b^{\pm} \rangle |v \rangle  -  \langle b^{PV} | g  |v \rangle |b^{\pm} \rangle  \right. \nonumber \\
   \left. + \langle b | g  |b^{PV \pm} \rangle |v \rangle  -  \langle b | g  |v \rangle |b^{PV \pm} \rangle \right. \nonumber \\
   \left. + \langle b^{PV \mp} | g  |b \rangle |v \rangle -  \langle b^{PV \mp} | g  |v \rangle |b \rangle  \right ] .
\label{eqhfu4}
\end{eqnarray}

\begin{figure}[t]
    \centering
    \includegraphics[scale=0.35]{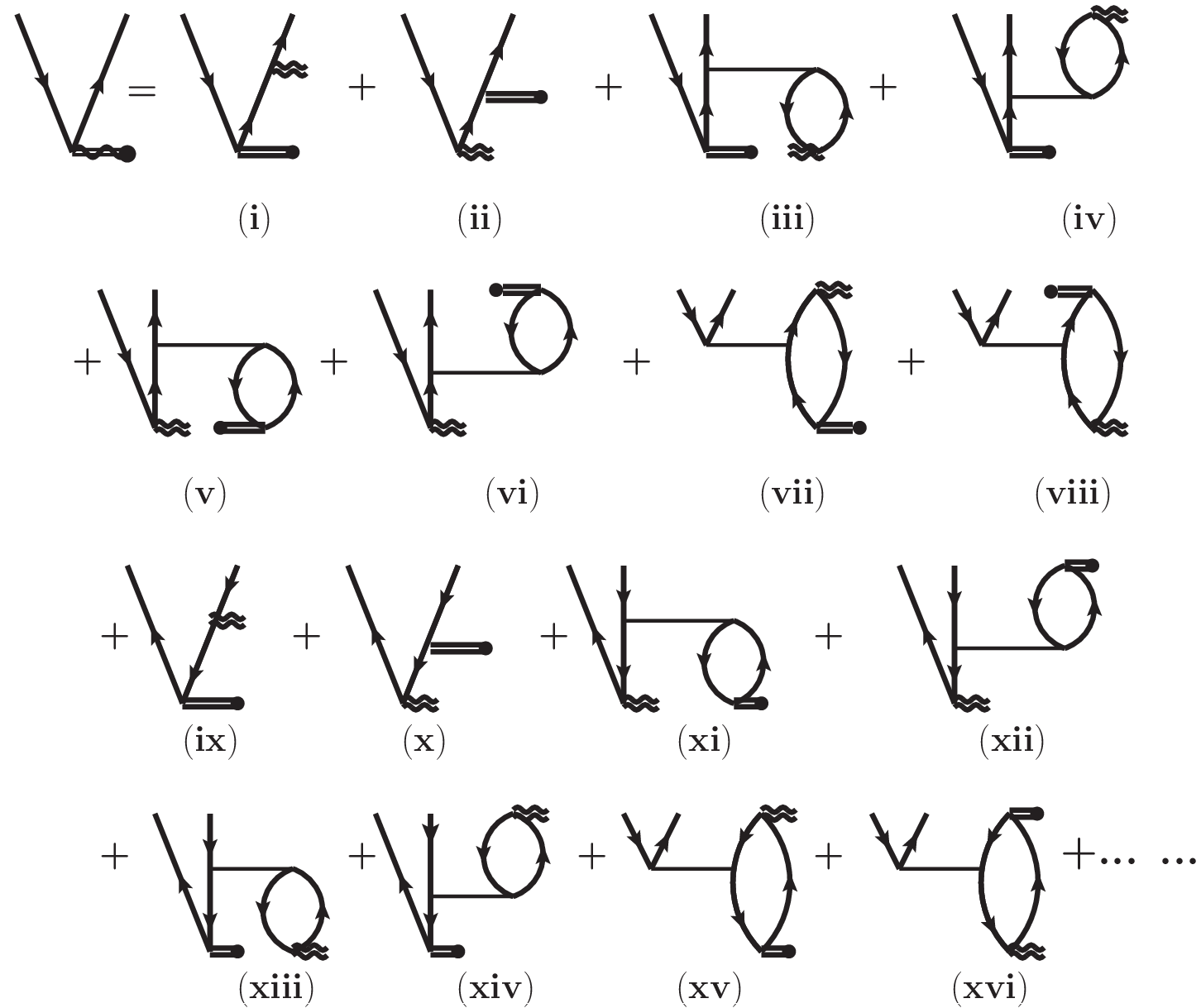}
    \caption{Goldstone diagrams representing terms of Eq. (\ref{eqhf4}) representing the $v^{PV\pm}$ term that gives rise to DCP contributions in the CPDF-RPA method. Diagrams from (ix) to (xvi) along with their exchanges are coming due to the implementation of the orthogonalization condition.}
    \label{CPHF-RPA}
\end{figure}

It can be further noted that in the CPDF method the perturbed core DHF orbital ($| a^{PV} \rangle $) is orthogonal to the unperturbed core orbital ($| a \rangle$) and the same is also true in the RPA. i.e. $\langle a | a^{PV} \rangle=0$ and $\langle a | a^{\pm} \rangle=0$. However, $\langle a| a^{PV \pm } \rangle \ne 0$ in the CPDF-RPA method. This necessitates to use the orthogonalized core orbitals ($| a^{o \pm} \rangle$) by imposing the condition  
\begin{eqnarray}
| a^{o \pm} \rangle &=& |a^{PV \pm } \rangle - \sum_b |b \rangle \langle b |  a^{PV \pm } \rangle .
\end{eqnarray}

In Fig. \ref{CPHF-RPA}, we show the Goldstone diagrams contributing to the determination of  $|a^{PV \pm } \rangle$ and also the diagrams that are subtracted to obtain $| a^{o \pm} \rangle$. It can be realized below that it is not required to obtain the modified orbitals for the virtual orbitals in the CPDF-RPA method to estimate the $E1_{\text{PV}}$ amplitude. 

Following the general formula given by Eq. (\ref{cprpa1}), we can write
\begin{eqnarray}
E1_{\text{PV}} &=& \langle f | d + u_i^+ | i^{PV} \rangle 
     +  \langle f | h_w + u_i^{PV} | i^+ \rangle \nonumber \\
 &&  + \langle f | u_i^{PV +} | i \rangle \nonumber \\
 &=& \langle \Phi_f | D \Omega^{i,CPDF} | \Phi_i \rangle + \langle \Phi_f | H_W \Omega^{i,+} | \Phi_i \rangle \nonumber \\
   && + \langle \Phi_f | \Omega^{CPDF + } | \Phi_i \rangle.
 \label{cprpa3}
\end{eqnarray}
Similarly, using the formula given by Eq. (\ref{cprpa2}) we can get
\begin{eqnarray}
E1_{\text{PV}} &=& \langle f^{PV} | d + u_i^+ | i \rangle 
     +  \langle f^{-} | h_w + u_i^{PV} | i \rangle \nonumber \\
 &&   + \langle f | u_f^{PV -} | i \rangle \nonumber \\
  &=& \langle f^{PV} | d + u_i^+ | i \rangle 
     +  \langle f^{-} | h_w + u_i^{PV} | i \rangle \nonumber \\
 &&   + \langle f | u_i^{PV +} | i \rangle \nonumber \\
 &=& \langle \Phi_f | \Omega^{f,CPDF \dagger} D | \Phi_i \rangle + \langle \Phi_f | \Omega^{f- \dagger} H_W | \Phi_i \rangle \nonumber \\
&& + \langle \Phi_f | \Omega^{f,CPDF -} | \Phi_i \rangle .
  \label{cprpa4}
\end{eqnarray}
In the above expressions, we define
\begin{eqnarray}
 \Omega^{CPDF + } = \sum_{i, j} ( \langle f | u_i^+ | i^{PV} \rangle +     \langle f | u_i^{PV} | i^+ \rangle  \nonumber \\ 
  + \langle f | u_i^{PV +} | i \rangle ) a_j^{\dagger} a_i 
\end{eqnarray}
and
\begin{eqnarray}
 \Omega^{CPDF - } = \sum_{i, j}  ( \langle f^{PV} | u_f^- | i \rangle +     \langle f^- | u_f^{PV} | i \rangle  \nonumber \\ 
   + \langle f | u_f^{PV -} | i \rangle  ) a_j^{\dagger} a_i .
\end{eqnarray}

\begin{figure}[t]
    \centering
    \includegraphics[height=80mm,width=80mm]{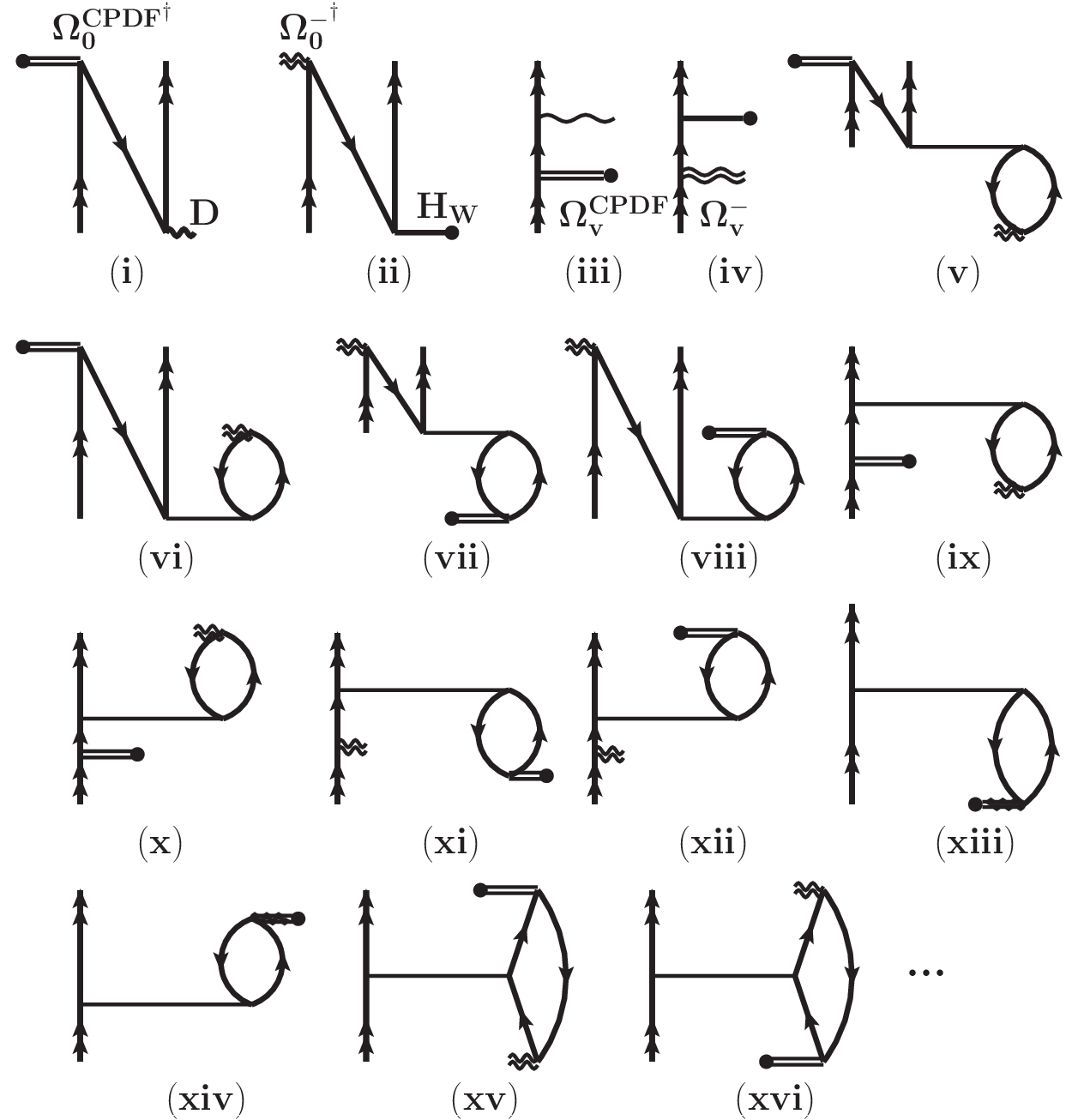}
    \caption{Goldstone diagrams contributing to the $E1_{\text{PV}}$ amplitude in the CPDF-RPA method.}
    \label{CPHF-RPA1}
\end{figure}

It is also worth noting that some of the works in the literature quoted contributions from $\langle f | u_i^{PV +} |i\rangle$ in the CPDF-RPA  method as `DCP' effects. In Fig. \ref{CPHF-RPA1}, we show the diagrams that are contributing to $E1_{\text{PV}}$ in the CPDF-RPA method including the DCP effect. Results without the DCP effects are mentioned as CPDF-RPA* method in order to compare them with the calculations reported in Ref. \cite{Dzuba2012}.

\begin{figure}[t]
    \centering
    \includegraphics[scale=0.40]{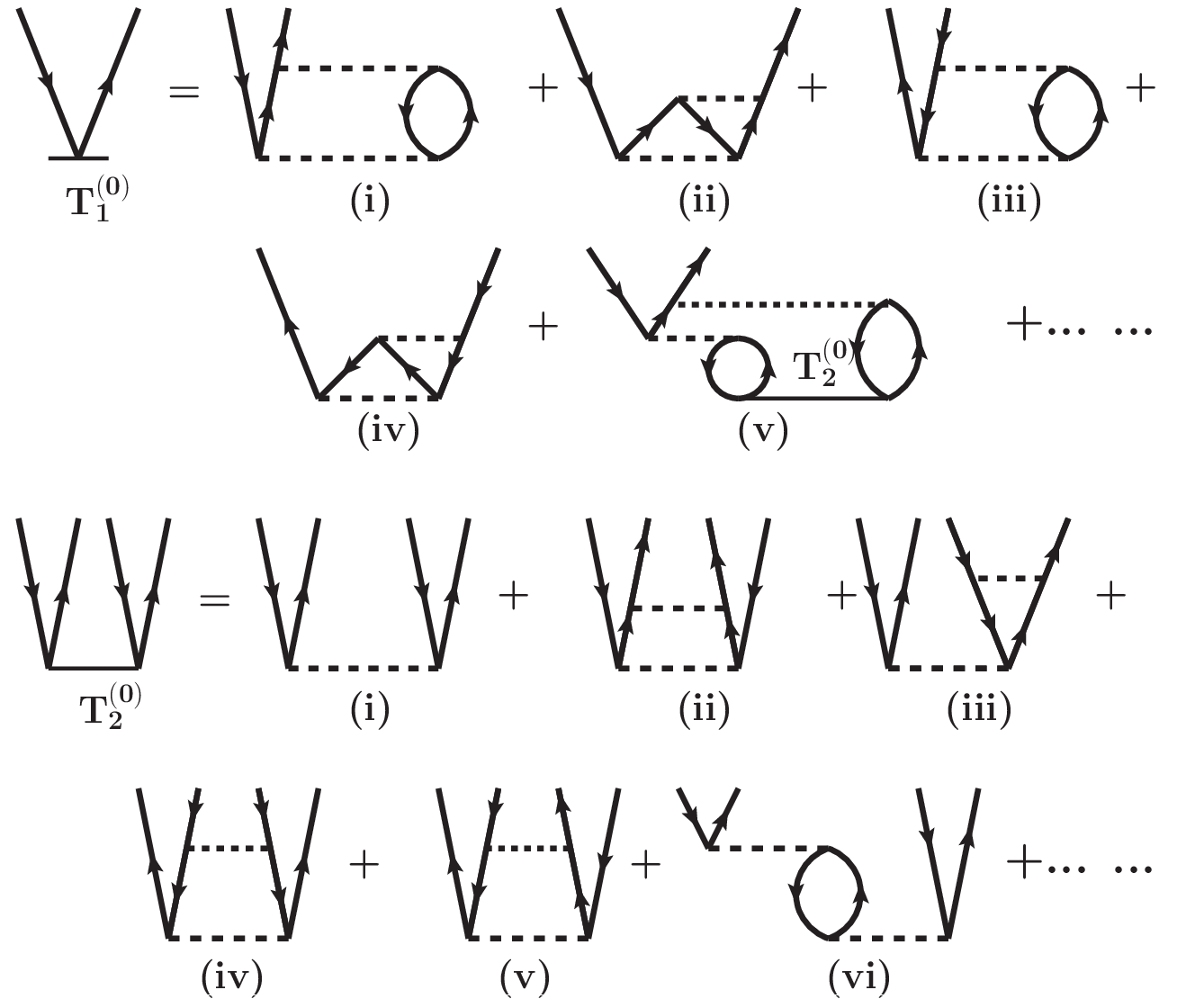}
    \caption{Goldstone diagrams showing breakdown of the $T^{(0)}$ operators in terms of lower-order perturbative excitations.}
    \label{RCC-T0}
\end{figure}

\begin{figure}[t]
    \centering
    \includegraphics[scale=0.4]{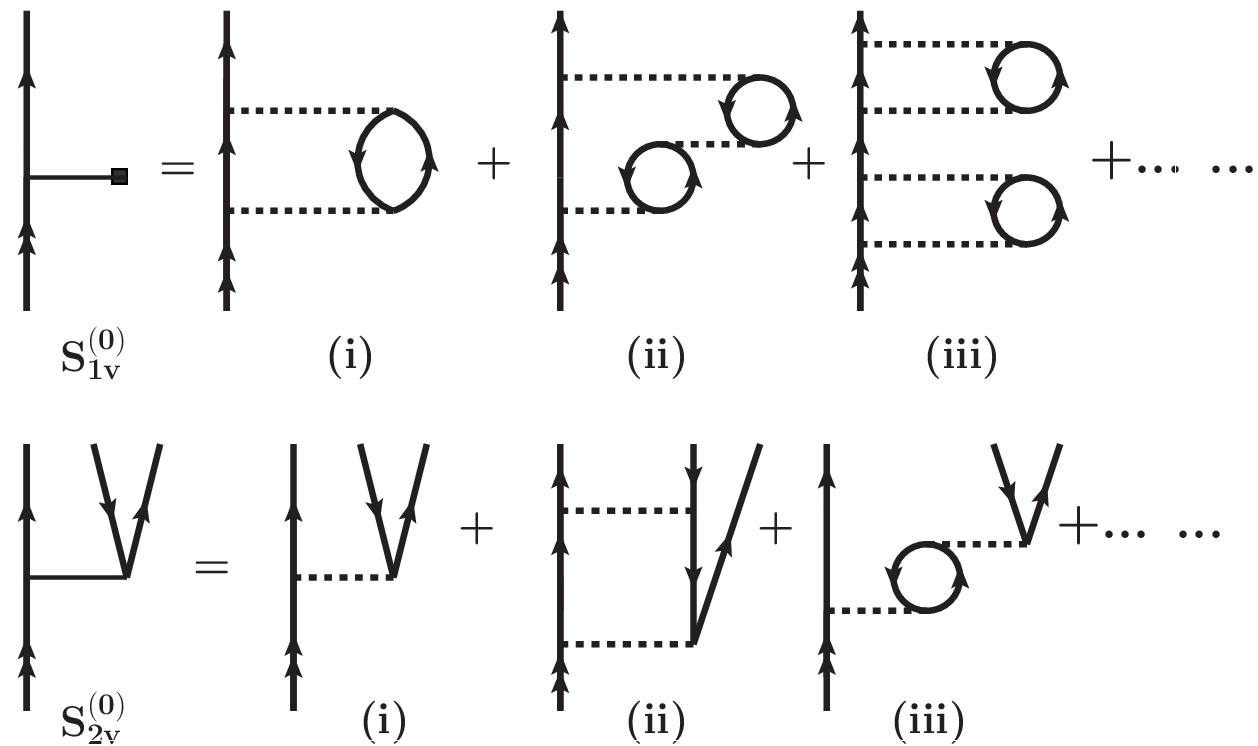}
    \caption{Goldstone diagrams showing breakdown of the $S_v^{(0)}$ operators in terms of lower-order perturbative excitations.}
    \label{RCC-Sv0}
\end{figure}

Starting with $|\Phi_v \rangle = a_v^{\dagger} |\Phi_0 \rangle$, we consider the multi-reference Fock-space approach in the RCC theory to define the exact wave function as \cite{Lindgren, Haque1984, Mukherjee1979},
\begin{eqnarray}
|\Psi_v \rangle = e^S |\Phi_v \rangle , 
\end{eqnarray}
where the RCC excitation operator due to correlation effects takes the form
\begin{eqnarray}
S = \sum_{k=0}^m \sum_{l=0}^n S^{(k,l)} =  S^{(0,0)} +  S^{(0,1)} +  S^{(1,0)} + \cdots. 
\end{eqnarray}
In this expression, $m$ and $n$ stand for number of hole and valence electrons with respect to $|\Phi_0 \rangle$. Since $^{133}$Cs possesses only one valence electron, the above series terminates at
\begin{eqnarray}
S =  S^{(0,0)} +  S^{(0,1)}.
\end{eqnarray}
For brevity, we use the notation $S^{(0,0)} \equiv T$ and $S^{(0,1)}=S_v$ to denote excitations of electrons from $|\Phi_0 \rangle$ and  $|\Phi_v \rangle$, respectively. Accordingly we can write
\begin{eqnarray}
|\Psi_v \rangle &=& e^{T+S_v} |\Phi_v \rangle = e^T \left \{ 1+ S_v \right \}  |\Phi_v \rangle .  
\end{eqnarray}
Here $e^{S_v} = 1+ S_v$ is the exact form for the one valence atomic systems. In the RCCSD method, the excitation operators are denoted as
\begin{eqnarray}
T = T_1 + T_2
\ \ \ \ 
\text{and} 
\ \ \ \
S_v = S_{1v} + S_{2v} ,
\end{eqnarray}
where subscripts 1 and 2 stand for the singles and doubles excitations respectively. In the wave operator form, it corresponds to
\begin{eqnarray}
\Omega_0 = e^T \ \ \ \ \text{and} \ \ \ \ \Omega_v = e^T S_v .
\end{eqnarray}

\begin{figure}[t]
    \centering
    \includegraphics[scale=0.4]{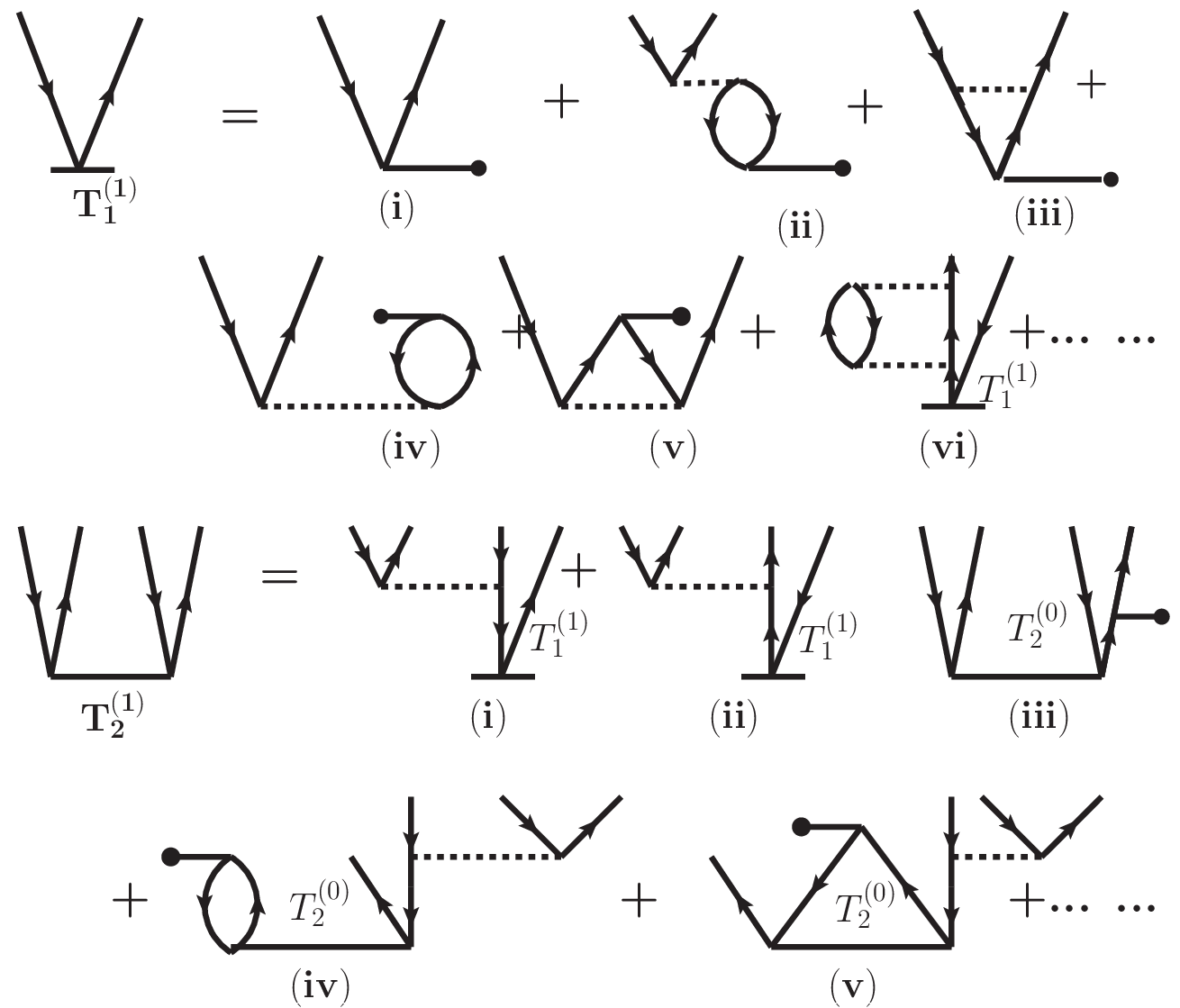}
    \caption{Goldstone diagrams showing breakdown of the $T^{(1)}$ operators in terms of lower-order perturbative excitations.}
    \label{RCC-T1}
\end{figure}

Following the Bloch's equations given by Eqs. (\ref{blw0}) and (\ref{blwv}), amplitudes of the $T$ and $S_v$ excitation operators are obtained by
\begin{eqnarray}
 \langle \Phi_0^* | (H e^T)_l | \Phi_0 \rangle =0 
\end{eqnarray}
and 
\begin{eqnarray}
 \langle \Phi_v^* | [(H e^T)_l - E_v] S_v  | \Phi_v \rangle = - \langle \Phi_v^* |  (H e^T)_l  | \Phi_v \rangle ,
 \label{eqsv1}
\end{eqnarray}
where subscript $l$ denotes for the linked terms, bra states with superscript $*$ means excited states with respect to the respective DHF ket states appear in the equations. In the {\it ab initio} procedure, the energy of the $|\Psi_v \rangle$ is determined by calculating expectation value of the effective Hamiltonian; i.e. 
\begin{eqnarray}
E_v&=&\langle \Phi_v|H_{eff}|\Phi_v \rangle = \langle \Phi_v| (H e^T )_l \left \{ 1 + S_v \right \} |\Phi_v \rangle . \ \ \
\label{heff1}
\end{eqnarray}
As can be noticed $E_v$ is a function of $S_v$ and $S_v$ itself depends on $E_v$. Thus, the non-linear Eqs. (\ref{eqsv1}) and (\ref{heff1}) are solved iteratively to obtain amplitudes of $S_v$. As pointed out earlier, appearance of $E_v$ in the determination of $S_v$ amplitudes is a consequence of using orbitals from $V^{N-1}$ potential. To evaluate $E1_{\text{PV}}$, we expand the $T$ and $S_v$ operators by treating $H_W$ as the perturbation to separate out the solutions for the unperturbed and the first-order wave functions by expressing
\begin{eqnarray}
T = T^{(0)} + \lambda_2 T^{(1)} 
\ \ \ \ 
\text{and} 
\ \ \ \
S_v = S_v^{(0)} + \lambda_2 S_v^{(1)} ,
\end{eqnarray}
where superscript meanings are same as specified earlier. 

\begin{figure}[t]
    \centering
    \includegraphics[scale=0.4]{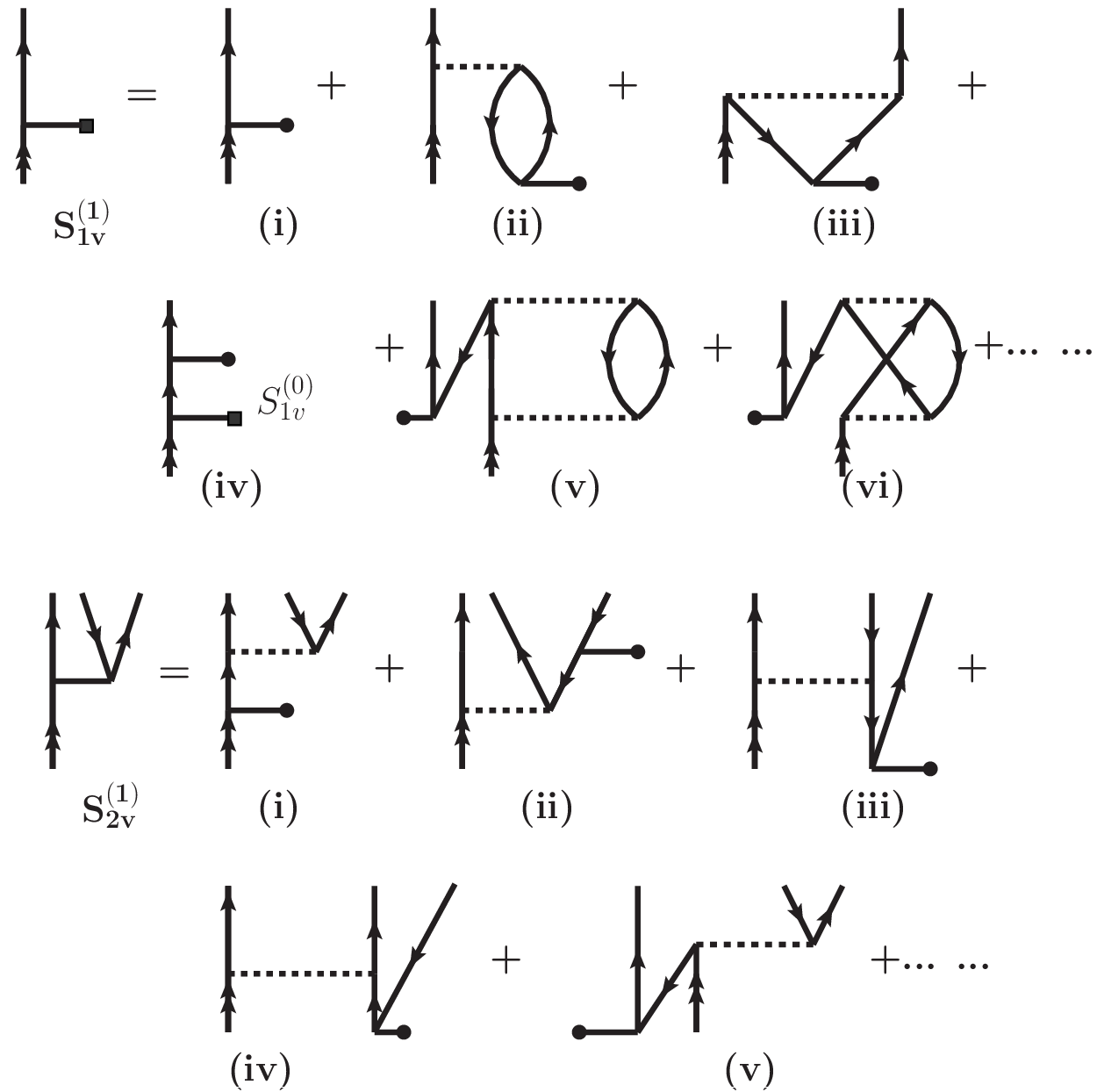}
    \caption{Goldstone diagrams showing breakdown of the $S_v^{(1)}$ operators in terms of lower-order perturbative excitations.}
    \label{RCC-Sv1}
\end{figure}

\begin{figure}[t]
    \centering
    \includegraphics[height=80mm,width=80mm]{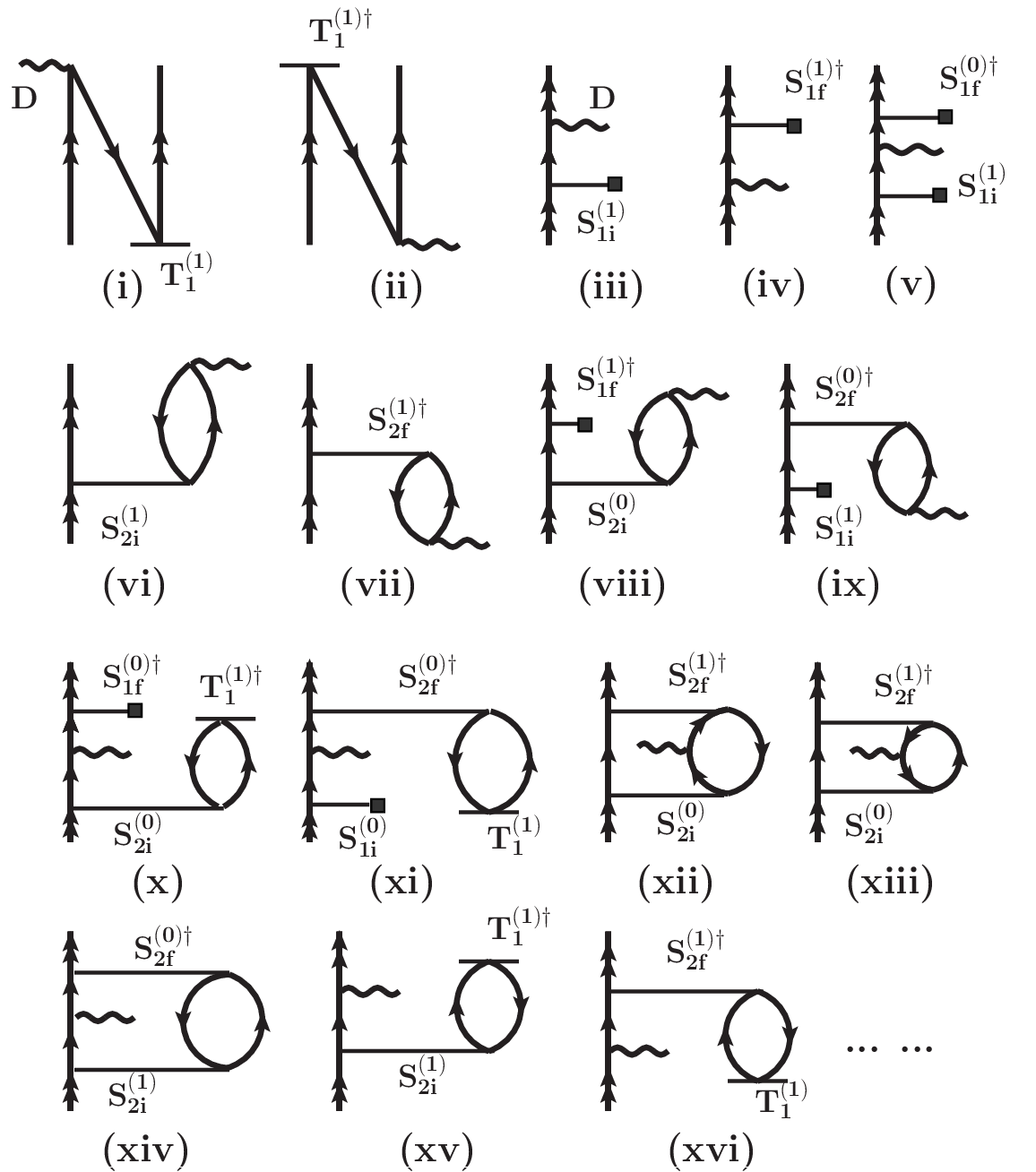}
    \caption{A few important $E1_{\text{PV}}$ evaluating diagrams in the RCCSD method. }
    \label{RCC-prop}
\end{figure}

The $E1_{\text{PV}}$ expression between the states $| \Psi_i \rangle$ and $| \Psi_f \rangle$ in the RCC theory is given by
\begin{eqnarray}
 E1_{\text{PV}} = \frac{\langle \Phi_f | \{S_f^{(1)\dagger} + (S_f^{(0)\dagger} +1) T^{(1)\dagger}\} \bar{D} \{ 1+ S_i^{(0)} \} |\Phi_i \rangle} {\langle \Phi_f | \{S_f^{(0)\dagger} +1 \} \bar{N} \{ 1+ S_i^{(0)} \} |\Phi_i \rangle} && \nonumber \\
 +  \frac{\langle \Phi_f |\{ S_f^{(0)\dagger} +1 \} \bar{D} \{T^{(1)}(1+ S_i^{(0)}) + S_i^{(1)}\} |\Phi_i \rangle}{\langle \Phi_f | \{S_f^{(0)\dagger} +1 \} \bar{N} \{ 1+ S_i^{(0)} \} |\Phi_i \rangle} , \ \ \ &&
\label{e1pnc}
\end{eqnarray}
where $\bar{D}=e^{T^{(0)+}}De^{T^{(0)}}$ and $\bar{N}=e^{T^{(0)+}}e^{T^{(0)}}$. Unlike the CPDF, RPA and CPDF-RPA methods, normalization factors appear explicitly in the RCC expression. Following Eqs. (\ref{eqcc}) and (\ref{eqvc}), one can easily identify RCC terms contributing to the Core and Valence correlations in the evaluation of $E1_{\text{PV}}$. It basically means any term connected either with the $S_{n=i,f}{^{(0/1)}}$ operators or with their conjugate operators is a part of the Valence correlation otherwise they will be a part of the Core correlation. It can be further clarified that the definitions of Core and Valence correlation contributions to $E1_{\text{PV}}$ in our RCC theory are in the line of the RMBPT$^w$ and CPDF methods as $H_W$ is considered as perturbation in our treatment. In Figs. \ref{RCC-T0},\ref{RCC-Sv0}, \ref{RCC-T1}, \ref{RCC-Sv1} we show some of the important Goldstone diagrams contributing to the $T_1^{(0)}$, $T_2^{(0)}$, $S_{1v}^{(0)}$, $S_{2v}^{(0)}$, $T_1^{(1)}$, $T_2^{(1)}$, $S_{1v}^{(1)}$ and $S_{2v}^{(1)}$ amplitudes. These diagrams can be compared with the CPDF-RPA wave operator amplitude determining diagrams to understand how they are embedded within the RCC operators. In addition, the RCC theory includes $E_v$ in the amplitude determining equation as should be the case for $V^{N-1}$ potential implying that it is correct representation of many-body theory. In Fig. \ref{RCC-prop} we show a few important contributing Goldstone diagrams from the RCC method to Core and Valence correlations. It can be followed that the RCC method includes correlation effects from core-polarization, pair-correlation, and DCP to all orders.

\begin{figure}[t]
    \centering
 \includegraphics[height=60mm,width=85mm]{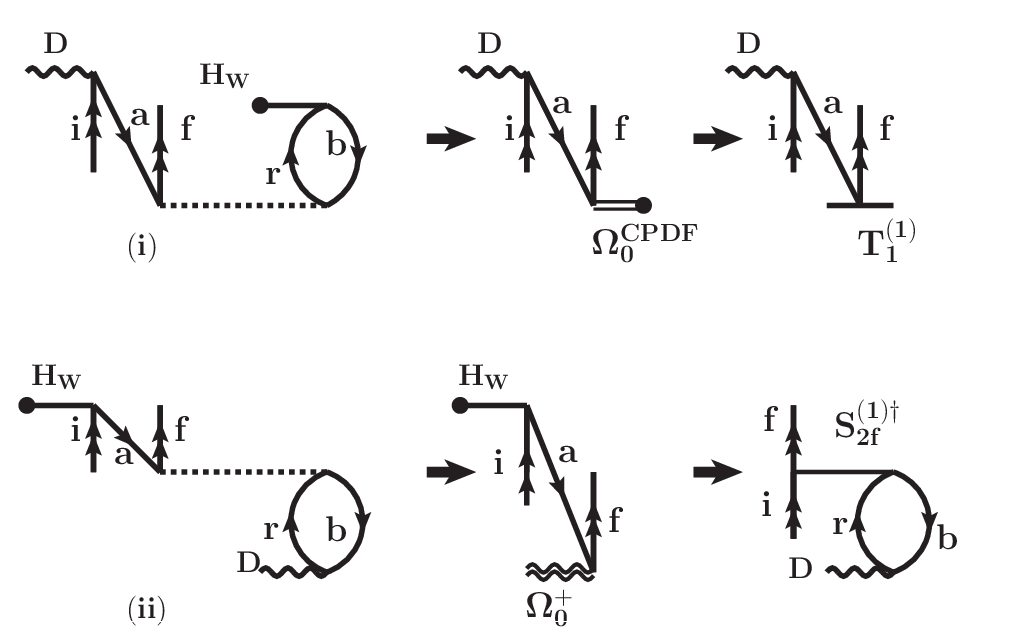}
\caption{Relating Core contributing diagrams from (i) $D \Omega_0^{(1,1)}$ of the RMBPT(3)$^w$ method and (ii) $H_W \tilde{\Omega}_0^{(1,1)}$ of the RMBPT(3)$^d$ method to the Core contributing diagrams in the CPDF-RPA method. The former remains as a Core contributing diagram while the later turns out to be a Valence contributing diagram in the RCCSD method.}
    \label{RCC-Sv2}
\end{figure}

At this juncture, we would like to demonstrate how some of the lower-order Core contributing diagrams of the RMBPT(3)$^w$ and RMBPT(3)$^d$ methods propagate to CPDF-RPA and RCCSD methods. For this purpose, we consider each representing diagram from the RMBPT(3)$^w$ and RMBPT(3)$^d$ methods as shown in Fig. \ref{RCC-Sv2}(i) and (ii), respectively. As can be seen from these figures, a Core contributing diagram from the $D \Omega_0^{(1,1)}$ term of the  RMBPT(3)$^w$ method propagates as $D \Omega_0^{CPDF}$ in the CPDF-RPA method and as $D T_1^{(1)}$ in the RCCSD method. Since $T_1^{(1)}$ contains both core-polarization and pair correlation effects to all-orders, the Core contributions estimated using the RCCSD method are more rigorous than the CPDF-RPA method. Returning to the Core contributing diagram from the $H_W \tilde{\Omega}_0^{(1,1)}$ term of the  RMBPT(3)$^d$ method, it becomes a part of the Core contributing term $H_W\Omega_0^+$ of the CPDF-RPA method. However, this is embedded within the $S_{2f}^{(1)\dagger} D$ term of the RCCSD method representing a part of the Valence correlation. This implies that the definitions of Core and Valence contributing terms in the considered CPDF-RPA and RCCSD methods are not unique. Additionally, the RCCSDT method used in Ref. \cite{Sahoo2021} is better than the RCCSD method under consideration, which implies that all Core contributions included in Ref. \cite{Dzuba2012} are also taken into account in Ref. \cite{Sahoo2021}.

\section{Results \& Discussions}

\begin{table}[t]
\caption{$E1_{\text{PV}}$ values, in units $10^{-11}i(-Q_w/Nn)|e|a_0$, of the $6s ~ ^2S_{1/2}-7s ~ ^2S_{1/2}$ and $6s ~ ^2S_{1/2}- 5d ~ ^2D_{3/2}$ transitions in $^{133}$Cs from DC Hamiltonian reported by various works. Methods shown as `Sum-over' and `Mixed' are obtained using sum-over-states approach and mixed many-body methods respectively. Results shown in bold fonts are claimed to be within 0.5\% accuracy.}
\begin{tabular}{lll ll }
    \hline \hline \\
   Method   & This work &  Others & This work & Others \\
         \hline \\
\multicolumn{3}{c}{\underline{$6s ~ ^2S_{1/2}-7s ~ ^2S_{1/2}$}} & \multicolumn{2}{c}{\underline{$6s ~ ^2S_{1/2}-5d ~ ^2D_{3/2}$}} \\
 & &  & & \\
DHF &  $0.7375$ & $0.736$ \cite{maartensson1985} & $-2.3933$ &  \\
RMBPT(3)$^w$  & $1.0902$& &$-2.4639$ \\
CPDF & $0.9226$  & $0.924$ \cite{maartensson1985}  & $-2.7989$  &   \\
RPA & $0.7094$ &$0.707$ \cite{maartensson1985}  & $-2.2362$ &   \\
CPDF-RPA$^*$ & $0.8876$  & $0.8914$ \cite{Dzuba2012}  & $-3.1665$  & $-3.80$ \cite{Roberts2013}   \\
             &            &  $0.8923^{\dagger}$ \cite{Dzuba2012} &  \\ 
           &    & $0.907$ \cite{Roberts2013} &  \\            
CPDF-RPA &$0.8859$ &  $0.886$ \cite{maartensson1985} & $-3.1071$ &$-3.70$ \cite{Roberts2013-2} \\
           &    & $0.9041$ \cite{Dzuba2002} &  \\ 
RCCSD & $0.8964$& $0.8961$ \cite{Sahoo2021} & $-3.5641$&$-3.210$\cite{Sahoosymm} \\
RCCSDT & &$\textbf{0.8967}$ \cite{Sahoo2021} &  \\
Sum-over & & $0.9053$ \cite{Porsev2010} &   & $-3.76$ \cite{Dzuba2001}  \\
         &  & $\textbf{0.8998}$$^{\dagger}$ \cite{Porsev2010} &  \\
Mixed-states & &$0.8967$ \cite{Dzuba2012} & $-3.62$ \cite{Dzuba2001} \\
      &  & 0.8938$^{\dagger}$ \cite{Dzuba2012}  &  &   \\
      &  & \textbf{0.9083}$^{\ddagger}$ \cite{Dzuba2012} &  \\
\hline \hline
\multicolumn{5}{l}{$^\dagger$Note: Scaled value.} \\ 
\multicolumn{5}{l}{$^\ddagger$Scaled value $+$ borrowed contribution from Ref. \cite{Porsev2010}.}
\end{tabular}
\label{tab1}
\end{table}

After explaining theoretically how Core contributions to $E1_{\text{PV}}$ can be different from one to another method, we intend to show this quantitatively in this section. We present calculated values of $E1_{\text{PV}}$ of the $6S-7S$ and $6S -5D_{3/2}$ transitions in $^{133}$Cs from the DHF, RMBPT(3), CPDF, RPA, CPDF-RPA and RCCSD methods. The main intention of the study is to demonstrate similarities/differences among various contributions to $E1_{\text{PV}}$, and to address the issue of the sign for the Core correlation contributions to the value. We have allowed correlations from all the occupied orbitals and excitations of electrons from a given set of virtual orbitals in all the considered many-body methods to make comparative analysis of results from them. We validate our calculations by comparing our values with the earlier reported values by M{\aa}rtensson \cite{maartensson1985}. The reason for presenting the $E1_{\text{PV}}$ value of the $6S-5D_{3/2}$ transition in $^{133}$Cs is to answer a Comment by Roberts and Ginges in Ref. \cite{Roberts2022}, where they argue about the agreement of the sign of Core contribution to the value of a $S-D$ transition reported using the RCCSD method \cite{Wansbeek2008}.

\begin{table}[t]
    \caption{Reduced E1 (in a.u.) and $H_W$ (in units of $10^{-11}i(-Q_w/Nn)|e|a_0$) matrix elements from the RCCSD method, and the excitation energies (in cm$^{-1}$) of the low-lying states of $^{133}$Cs that are used to estimate the `Main' contribution of $E1_{\text{PV}}$ for the $6S-7S$ transition. E1 matrix elements and energies are compared with the available most precise experimental values.}
        \begin{tabular}{l c c c c c}
        \hline\hline\\
Transition & \multicolumn{2}{c}{E1 amplitude}& \multicolumn{2}{c}{Excitation Energy} & $H_W$ amplitude \\
& \cline{1-4}\\
& This work & Experiment &This work& Experiment \cite{NIST} \\ 
\hline
$6P_{1/2}$-$6S$ & $4.5487$ & $4.5097(74)$ \cite{Young1994} & $-11243.93$ & $-11178.27$ & $-1.2541$\\
$7P_{1/2}$-$6S$ & $0.3006$ & $0.2825(20)$ \cite{Vasilyev2002} & $-21838.93$ & $-21765.35$ & $-0.7135$\\
$8P_{1/2}$-$6S$ & $0.0914$ & & $-25787.48$ & $-25708.83$ & $-0.4808$\\
$9P_{1/2}$-$6S$ & $-0.0388$ & & $-27735.96$ & $-27637.00$ & $0.3471$\\
$6P_{1/2}$-$7S$ & $-4.2500$ &$4.233(22)$\cite{Bouchiat1984} & $7352.53$ & $7357.26$ & $0.6067$\\
$7P_{1/2}$-$7S$ & $10.2967$ & $10.308(15)$\cite{Bennett1999} & $-3242.47$ & $-3229.82$ & $0.3445$\\
$8P_{1/2}$-$7S$ & $0.9492$ & & $-7191.02$ & $-7173.31$ & $0.2320$\\
$9P_{1/2}$-$7S$ & $-0.3867$ & & $-9139.50$ & $-9101.47$ & $-0.1674$\\
\hline\hline
        \end{tabular}
\label{tab2}
\end{table}

Table \ref{tab1} presents the $E1_{\text{PV}}$ values of the aforementioned transitions, in units of $10^{-11}i(-Q_w/Nn)|e|a_0$ with neutron number $Nn$, using the DC Hamiltonian from different methods. The values shown in bold fonts in this table are claimed to be accurate within 0.5\% by earlier works, but some of them differ by 1\% from each other. The sum-over-states approach, given as `Sum-over' in the table, uses scaled E1 matrix elements and energies from the CCSDvT method to estimate the Main contribution, while the X-factor is obtained using a blend of many-body methods \cite{Porsev2010}. In Refs. \cite{Dzuba2012}, the same `Main' contribution is utilized but Core and Tail contributions to the X-factor are estimated using the CPDF-RPA* method (denoted as only RPA in the original paper). Pair-correlation effects to these estimations were estimated using the BO-correlation method. Result from these RPA$+$BO methods is given under `Mixed-states' in the above table. The large discrepancy seen in both the results from the above table come from the X-factors estimated in Refs. \cite{Porsev2010, Dzuba2012}. This could be attributed to distribution of contributions under the Core and Valence correlations in the considered approaches in both the works. However, comparison with the RCCSD method does not support such distributions.

\begin{table}[t]
\caption{Estimated `Main' contributions to the $E1_{\text{PV}}$ values, in units $10^{-11}i(-Q_w/Nn)|e|a_0$, of the $6s ~ ^2S_{1/2}-7s ~ ^2S_{1/2}$ transition in $^{133}$Cs using matrix elements involving $np ~ ^2P_{1/2}$ intermediate states in the sum-over-states approach. Four cases are being considered: (a) $\textit{ab initio}$ result in which calculated values from Table \ref{tab2} are used; (b) replacing calculated E1 matrix elements by their experimental values;  (c) retaining calculated E1 matrix elements and using experimental energies; and
(d) using experimental values for both E1 matrix elements and energies.}

 \begin{tabular}{c c c c}
 \hline\hline\\
Approach & $\langle7S|D|6S^{PNC}\rangle$ & $\langle 7S^{PNC}|D|6S\rangle$ & Total \\
\hline
(a) & $-0.4461$ & $1.3171$ & $0.8710$\\
(b) & $-0.4373$ & $1.3121$ & $0.8748$\\
(c) & $-0.4522$ & $1.3156$ & $0.8634$\\
(d) & $-0.4434$ & $1.3106$ & $0.8672$\\
\hline\hline
\end{tabular}
\label{tab3}
\end{table}

To understand the reasons for significant discrepancies seen in the X-factors from various works, we analyze the Main contribution to the $6S-7S$ transition by using the calculated properties from our RCCSD method in the sum-over-states approach. We present the $H_W$ matrix elements, E1 matrix elements and energies obtained using the DC Hamiltonian in Table \ref{tab2} , and compare them with experimental values \cite{Young1994, Vasilyev2002, Bouchiat1984, Bennett1999, NIST}. Using these values we estimate the Main contributions to $E1_{\text{PV}}$ of the $6S-7S$ transition and they are given in Table \ref{tab3}. Results (a) from {\it ab initio} calculations, (b) using experimental E1 values with calculated energies, (c) calculated E1 values with experimental energies and (d) using experimental values for both the E1 matrix elements and energies are given separately. This analysis suggests that accuracy of energies in a given method affect the results more than accuracy of matrix elements because value from (b) is larger than (a), but values from (c) and (d) are lower than (a). Differences between the {\it ab initio} and semi-empirical calculations can be minimised by including contributions from the triples, quadruples etc. higher-level excitations of the RCC method. In Ref. \cite{Sahoo2021}, Sahoo et al. have demonstrated that the difference between the {\it ab initio} calculations of $E1_{\text{PV}}$ of the $6S-7S$ transition using the RCCSD and RCCSDT methods is very small. Core contributions are almost the same in both the methods, and the agreement of the values from both the methods was the result of opposite trends of correlation effects in the evaluation of the $H_W$ matrix elements than the E1 matrix elements and energies. Subtracting the {\it ab initio} value of Main from the final RCCSD result, we find the X-factor to $E1_{\text{PV}}$ of the $6S-7S$ transition as 0.0254, against 0.0175 and 0.0256 of Refs. \cite{Porsev2010} and \cite{Dzuba2012} respectively in units of $10^{-11}i(-Q_w/Nn)|e|a_0$. It means that there is a large difference between the X-factor of Ref. \cite{Porsev2010} and our work, whereas these values almost agree between Ref. \cite{Dzuba2012} and the present work. Since there is a sign difference between the Core contribution from Ref. \cite{Dzuba2012} and the RCCSD value of Ref. \cite{Sahoo2021}, the above analysis suggests that the sign difference is solely due to different definitions used for the Core contribution in both the works.

\begin{table}[t]
   \caption{Core and Valence correlation contributions to 
  the $E1_{\text{PV}}$ values, in units $10^{-11}i(-Q_w/Nn)|e|a_0$, for the $6s ~ ^2S_{1/2}-7s ~ ^2S_{1/2}$ and $6s ~ ^2S_{1/2}- 5d ~ ^2D_{3/2}$ transitions in $^{133}$Cs from the RMBPT(3)$^w$ and RMBPT(3)$^d$ approaches. Results (a) considering $H_{eff}$ effect due to $V^{N-1}$ potential and (b) using DHF orbital energies are shown for comparison.}
    \begin{tabular}{lcccc}
    \hline\hline\\
  Approach&\multicolumn{2}{c}{RMBPT(3)$^w$}&\multicolumn{2}{c}{RMBPT(3)$^d$}\\
  \cline{2-5}
  &Core&Valence&Core&Valence\\ 
  \hline
  & &\multicolumn{2}{c}{{\underline{$6s ~ ^2S_{1/2}-7s ~ ^2S_{1/2}$}}}&\\
     (a)&$-0.00205$&$1.09220$&$-0.00003$&$1.24290$\\
     (b)&$-0.00206$&$0.43938$&$-0.00031$&$0.43763$\\
     \hline\hline\\
  & &\multicolumn{2}{c}{\underline{$6s ~ ^2S_{1/2}-5d ~ ^2D_{3/2}$}}&\\  
  (a)&$-0.16391$&$-2.30000$&$-0.12208$&$-2.55427$\\   
  (b)&$-0.18267$&$-3.97004$&$-0.12513$&$-4.02758$\\
  \hline\hline
    \end{tabular}   
    \label{RMBPT3}
\end{table}

To explain how definition of Core contribution changes depending upon the choice of an approach to estimate $E1_{\text{PV}}$, we present the Core and Valence contributions separately to the $6S-7S$ and $6D-5D_{3/2}$ transitions from both the RMBPT(3)$^w$ and RMBPT(3)$^d$ approaches in Table \ref{RMBPT3}. To demonstrate how appearance of $H_{eff}$ in the wave function determining equation due to choice of $V^{N-1}$ modify the result, we present RMBPT(3) results considering effect of $H_{eff}$ (given results as (a) in the above table) and replacing it with DHF energy as the case of the CPDF-RPA method (corresponding results are given under (b)).  Core and Valence results from both approaches are coming out differently, but the final results from both methods are almost close to each other. Changes in the results for both the transitions are noticed to be enormous when $H_{eff}$ is considered.

\begin{table}[t]
\caption{{\it Ab initio} contributions to Core and Valence parts of the $E1_{\text{PV}}$ values, in units $10^{-11}i(-Q_w/Nn)|e|a_0$, for the $6s ~ ^2S_{1/2}-7s ~ ^2S_{1/2}$ and $6s ~ ^2S_{1/2}- 5d ~ ^2D_{3/2}$ transitions in $^{133}$Cs from different methods considered in this work using the DC Hamiltonian. Available results from previous calculations are also given for comparison.}
\begin{tabular}{lcc cc}
    \hline \hline \\
   Method   & \multicolumn{2}{c}{This work} & \multicolumn{2}{c}{Others} \\
   \cline{2-3} \cline{4-5} \\
       & Core & Valence & Core & Valence \\
\hline \\
\multicolumn{5}{c}{\underline{$6s ~ ^2S_{1/2}-7s ~ ^2S_{1/2}$}} \\
 & &  \\
DHF &  $-0.00173$ & $0.73923$  &$-0.00174$\cite{Roberts2022}  \\
RMBPT(3)$^w$  & $-0.00205$&$1.09220$ \\
RMBPT(3)$^d$  &$-0.00003$ &$1.24290$ \\
CPDF & $-0.00199$  & $0.92454$ &$-0.00201$ \cite{Roberts2022}  \\
RPA & $0.00028$ &$0.70912$  \\
CPDF-RPA$^*$ & $0.00169$  & $0.88591$  & $0.00170$ \cite{Roberts2022} \\
CPDF-RPA & $0.00169$ & $0.88421$ & & \\
RCCSD & $-0.00197$& $0.89840$ & $-0.0019$ \cite{Sahoo2021} & $0.8980$ \cite{Sahoo2021}\\
\hline \\
\multicolumn{5}{c}{\underline{$6s ~ ^2S_{1/2}-5d ~ ^2D_{3/2}$}} \\
  & &  \\
DHF &  $-0.11684$ & $-2.27646$   \\
RMBPT(3)$^w$  &$-0.16391$&$-2.3000$ \\
RMBPT(3)$^d$  & $-0.12208$&$-2.55427$ \\
CPDF & $-0.19122$  &$-2.60768$  \\
RPA & $-0.12037$ &$-2.11585$  \\
CPDF-RPA$^*$ & $-0.20786$  & $-2.95860$  \\ 
CPDF-RPA  &$-0.20786$  &$-2.89923$  \\
RCCSD & $-0.14745$& $-3.41667$  & & \\
\hline \hline
\end{tabular}
\label{tab4}
\end{table}

\begin{table}[t]
\caption{Comparison of contributions from the initial and final perturbed states to $E1_{\text{PV}}$ of the $6s ~ ^2S_{1/2}-7s ~ ^2S_{1/2}$ transition of $^{133}$Cs, in units $10^{-11}i(-Q_w/Nn)|e|a_0$, at different levels of approximation between the present work and that are reported in Refs. \cite{maartensson1985,Roberts2022}.}
\begin{tabular}{l  cc  cc}
\hline\hline \\
Method & \multicolumn{2}{c}{$\langle{7S}^{PNC}|D|6S\rangle$} &  \multicolumn{2}{c}{$\langle7S|D|6S^{PNC}\rangle$} \\
\cline{2-3} \cline{4-5} \\
& Ours & Ref. \cite{maartensson1985} & Ours & Ref. \cite{maartensson1985} \\
\hline \\
& \multicolumn{4}{c}{\underline{Total contribution}} \\
DHF &$1.01168$&$1.010$&$-0.27418$&$-0.274$ \\
CPDF & $1.26664$&$1.267$&$-0.34409$&$-0.344$  \\
RPA  &$1.02557$&$1.023$&$-0.31617$&$-0.316$ \\
CPDF-RPA* &$1.27910$&$1.279$&$-0.39150$&$-0.391$ \\
\hline \\
       & Ours & Ref. \cite{Roberts2022} & Ours & Ref. \cite{Roberts2022} \\
\hline \\
& \multicolumn{4}{c}{\underline{Core contribution}} \\
DHF & $-0.02638$ & $-0.02645$ & $0.02465$ & $0.02472$ \\
CPDF & $-0.04298$ & $-0.04319$ & $0.04099$ & $0.04119$ \\
RPA  & $-0.03536$ &  & $0.03564$ &  \\
CPDF-RPA* & $ -0.05794$& $-0.05822$ & $0.05963$ & $0.05992$ \\
 & & & & \\
\multicolumn{5}{c}{\underline{Valence contribution}} \\
DHF &$1.03806$& &$-0.29883$& \\
CPDF & $1.30962$& & $-0.38508$& \\
RPA  & $1.06094$& &$0.70912$& \\
CPDF-RPA* & $1.33704$& &$-0.45113$& \\
\hline \hline
\end{tabular} 
\label{tab5}
\end{table}

To figure out about mismatch in the X-factors from various works, we present  the Core and Valence contributions separately for both the $6S-7S$ and $6D-5D_{3/2}$ transitions arising through the DHF method and many-body methods at a given level of approximation employed by different groups in Table \ref{tab4}. In Table \ref{tab5}, we compare the values reported in a Comment by Roberts and Ginges \cite{Roberts2022} and M{\aa}rtensson \cite{maartensson1985}. The results show that there is no issue with the implementation of these theories in our code. However, definitions of Core correlation effects arising through the CPDF, RPA and CPDF-RPA methods all differ. The exact reason for which sign of Core contribution to the $E1_{\text{PV}}$ of the transition in $^{133}$Cs differ between Ref. \cite{Porsev2010} and Ref.\cite{Dzuba2012} is not clear as the method(s) employed in Ref. \cite{Porsev2010} for its estimation is not mentioned explicitly. From the sign of the Core contribution we can assume that  Ref. \cite{Porsev2010} estimates Core contribution by considering $H_W$ as perturbation. However, large differences between the X-factors from Refs. \cite{Porsev2010}, \cite{Dzuba2012} and this work, which report as 0.0175, 0.0256 and 0.0254 in units of $10^{-11}i(-Q_w/Nn)|e|a_0$, suggest that the former work underestimates the Tail contribution. These differences are due to different levels of approximation in the many-body methods employed for their estimations.

Now, we wish to address the reason why Roberts and Ginges were able to get same sign for the Core contribution to $E1_{\text{PV}}$ of the $7S-6D_{3/2}$ transition in Ra$^+$ using their RPA$+$BO method with that are reported using the RCC method in Ref. \cite{Wansbeek2008}. Since correlation trends to $E1_{\text{PV}}$ of the $S-D_{3/2}$ transitions are almost similar in Cs and Ra$^+$, we can understand the above point by analysing the Core contributions to $E1_{\text{PV}}$ of the $6S-5D_{3/2}$ transition from different methods and comparing their trends with the $6S-7S$ transition of $^{133}$Cs.  Table \ref{tab4} shows that difference between the Core contributions from the RMBPT(3)$^{w}$ and RMBPT(3)$^d$ methods in the $6S-5D_{3/2}$ transition are small, and there is no sign difference between the CPDF and RPA results. These trends can be explained as follows. In the $6S-7S$ transition, wave functions of both the associated states have large overlap over the nucleus while in the $6S-5D_{3/2}$ transition only the wave function of the ground state has large overlap with the nucleus. As a result, strong core-polarization effects contribute through both the states in the former case.  Since core-polarization effects arising through the $D$ operator are stronger and have opposite signs than that arise through $H_W$, the net Core contributions in the $S-S$ and $S-D$ transitions behave very differently in the CDHF method and RPA, and the same propagates to the CPDF-RPA*/CPDF-RPA method. Since Core and Valence contributions are basically redistributed in the CPDF-RPA* and RCCSD methods, difference between the final values between Refs. \cite{Dzuba2012} and \cite{Sahoo2021} as well as from the present work are coming out to be very small in the $6S-7S$ transition, while it is not prominently noticeable in the $6S-5D_{3/2}$ transition (refer to Table \ref{tab1} for the comparison of results from the Mixed-states and RCCSD methods) . 

\begin{table}[t!]
\caption{Contributions to Core and Valence parts from different terms to $E1_{\text{PV}}$ of the $6s ~ ^2S_{1/2}-7s ~ ^2S_{1/2}$ transition in $^{133}$Cs, in units $10^{-11}i(-Q_w/Nn)|e|a_0$ from Eqs. (\ref{cprpa3}) and (\ref{cprpa4}) of the CPDF-RPA method, which are quoted under `Expression a' and `Expression b' respectively. Results are given using the calculated $\omega$ value, $\omega^{ex}$ and $\omega^{ex}$ with the experimental energies of the initial and final states (denoted by $E_{i,f}^{ex}$).} \label{tab6}
\begin{tabular}{lccc |ccc}
\hline \hline
Contribution   & \multicolumn{3}{c|}{Expression a}  & \multicolumn{3}{c}{Expression b} \\ 
\cline{2-4} \cline{5-7} \\
& \multicolumn{6}{c}{\underline{$6s ~ ^2S_{1/2}-7s ~ ^2S_{1/2}$}} \\
& $\langle7s|h_w|6s^+\rangle$ & $\langle7s|u_{6s}^{PV}|6s^+\rangle$ & Total & 
 $\langle7s^{PV}|d|6s\rangle$ & $\langle7s^{PV}|u_{6s}^{+}|6s\rangle$ & Total \\
 \hline \\
 Core ($\omega$) &$-0.0357$&$-0.02257$  &$-0.05794$ &  $-0.04299$&$-0.01495$ &$-0.05794$ \\
 Valence ($\omega$) &$1.06094$ &$0.27610$ &$1.33704$ & $1.30962$&$0.02742$ &$1.33704$\\
& & & & & & \\ 
 Core ($\omega^{ex}$) &$-0.03464$&$-0.02211$&$-0.05675$&$-0.04299$&$-0.01458$&$-0.05757$\\
 Valence ($\omega^{ex}$)&$-0.19464$&$-0.04598$&$-0.24062$&$1.30962$&$0.02743$&$1.33705$\\  
 & & & & & & \\ 
 Core ($E_{i,f}^{ex}$)&$-0.03546$&$-0.02283$&$-0.05829$&$-0.04331$&$-0.01498$&$-0.05829$\\
Valence ($E_{i,f}^{ex}$)&$1.21721$&$0.31956$&$1.53677$&$1.53384$&$0.00293$&$1.53677$\\
 \hline \\
& $\langle 7s|d|6s^{PV}\rangle$ & $\langle 7s|u_{6s}^{+}|6s^{PV}\rangle$ & Total & $\langle 7s^{-}|h_w|6s\rangle$ & $\langle7s^{-}|u_{6s}^{PV}|6s\rangle$ & Total\\
\hline \\
Core ($\omega$) &$0.04099$ &$0.01864$ &$0.05963$ &$0.03564$ &$0.023399$ &$0.05963$ \\
Valence ($\omega$) &$-0.38508$ &$-0.06605$ &$-0.45113$ &$-0.35181$ &$-0.09932$ &$-0.45113$ \\
& & & & & & \\
 Core ($\omega^{ex}$) &$0.04099$&$0.01915$&$0.06014$&$0.03651$&$0.02458$&$0.06109$\\
 Valence ($\omega^{ex}$)&$-0.38508$&$-0.06644$&$-0.45152$&$-0.17081$&$-0.05022$&$-0.22103$\\ 
 & & & & & & \\
Core ($E_{i,f}^{ex}$) &$0.04210$&$0.01999$&$0.06209$&$0.03686$&$0.02523$&$0.06209$\\
Valence ($E_{i,f}^{ex}$) &$-0.12128$&$-0.05800$&$-0.17928$&$-0.13743$&$-0.04185$&$-0.17928$\\
 \hline \hline\\
\end{tabular}
\end{table}

The DCP contributions from our calculations can be estimated by taking the differences in the results from the CPDF-RPA* and CPDF-RPA methods. This difference for the $6S-7S$ transition from our work is compared with the corresponding values from Refs. \cite{maartensson1985} and \cite{Roberts2013}. We find that our result agrees better with M{\aa}rtensson than Roberts. Also, our final CPDF-RPA result agrees well with Ref. \cite{maartensson1985} than Ref. \cite{Dzuba2002, Roberts2013}. We also intend to mention that the CPDF-RPA* results in Refs. \cite{Dzuba2012} and \cite{Roberts2022} are scaled by using $\omega^{ex}=0.0844$ a.u.. In the previous section, we have mentioned on the basis of theoretical analysis why such an approach would lead to errors in the determination of the $E1_{\text{PV}}$ values. To demonstrate it numerically, we have given results for the $6S-7S$ transition from the CPDF-RPA method using Eqs. (\ref{cprpa3}) and (\ref{cprpa4}) in Table \ref{tab6}. We have given these values using $\omega$, $\omega^{ex}$ and then also using $\omega^{ex}$ and experimental energies (denoted by $E_{i,f}^{ex}$) of the $6S$ and $7S$ states. From the comparison of the results, we observe a very a interesting trend. When both $\omega$ and energies of the atomic states are considered either from theory or experiment, results from both Eqs. (\ref{cprpa3}) and (\ref{cprpa4}) match each other, otherwise large discrepancies are seen. Nevertheless, it can be found from Table \ref{tab6} that our result with $\omega^{ex}$ value from the CPDF-RPA* method does not match with the corresponding results from Refs. \cite{Dzuba2012,Roberts2022} for the $6S-7S$ transition. We are unable to understand the reason for this though results with theoretical $\omega$ value from both the works agree quite well.

\begin{table}[t]
\caption{First-principle calculated $E1_{\text{PV}}$ values (in $-i (Q_{W}/Nn) ea_0 \times 10^{-11}$) of the $6s ~ ^2S_{1/2} - 7s ~ ^2S_{1/2}$ and $6s ~ ^2S_{1/2} - 5d ~ ^2D_{3/2}$ transitions in $^{133}$Cs from different terms of the RCCSD method. Here, contributions under `Norm' represents the difference between the contributions after and before normalizing the RCCSD wave functions. `Others' denote contributions from those RCCSD terms that are not shown explicitly in this table.}
\begin{tabular}{lrr}
\hline\hline\\
 RCC term & {$6s ^2S_{1/2} - 7s ~ ^2S_{1/2}$} &{$6s ~ ^2S_{1/2} - 5d ~ ^2D_{3/2}$}  \\
 \hline \\
    &  {\it Ab initio} & {\it Ab initio}  \\
\hline \\
 & \multicolumn{2}{c}{Core contribution} \\
 $\overline{D}T_1^{(1)}$  &$-0.04161$ &$-0.00062$\\
 $T_1^{(1)\dagger} \overline{D}$  & $0.03964$&$-0.17132$ \\
 Others  &$-0.00005$  &$0.01757$\\ 
Norm&$0.00005$&$0.00692$\\
 \hline  \\
& \multicolumn{2}{c}{Valence contribution} \\
 $\overline{D}S_{1i}^{(1)}$   &$-0.19363$&$-2.96310$\\
 $S_{1f}^{(1)\dagger} \overline{D}$     &$1.80382$ &$-0.89993$\\
 $S_{1f}^{(0)\dagger} \overline{D} S_{1i}^{(1)}$  &$-0.23184$&$-0.06863$\\
 $S_{1f}^{(1)\dagger} \overline{D} S_{1i}^{(0)}$  &  $-0.41826$&$0.10487$\\
 $\overline{D}S_{2i}^{(1)}$    &$-0.00039$&$0.00107$ \\
 $S_{2f}^{(1)\dagger} \overline{D}$   &  $0.00033$&$-0.00023$\\
 Others        & $-0.04040$&$0.24888$\\
 Norm& $-0.02122$& $0.16040$\\
 \hline\\
 Total &$0.89643$&$-3.56412$\\
 \hline\hline
  \end{tabular}
\label{tab7}
\end{table}

As mentioned earlier, three different approaches can be adopted  for evaluation of the $E1_{\text{PV}}$ amplitudes. The same applies to the RCC theory as well. However, we adopt the approach of evaluating the matrix element of the $D$ operator after considering $H_W$ as the external perturbation. Though this approach is in the line with the CPDF method, it is effectively takes care of electron correlation effects through both the $H_W$ and $D$ operators as in the CPDF-RPA method. In fact, it goes much beyond the CPDF-RPA method to include the electronic correlation effects which will be evident from the follow-up discussions. In this sense the RCCSDT method employed by Sahoo et al. \cite{Sahoo2021, SahooaRxiv} to estimate the $E1_{\text{PV}}$ amplitude of the $6S-7S$ transition in $^{133}$Cs includes the RPA contributions that are mentioned in Refs. \cite{Dzuba2012, Roberts2022}. However, some of these contributions are not a part of the Core contribution rather they come through the Valence contribution in our RCCSD method owing to the fact that the $D$ operator is not treated as an external perturbation here. This point can be comprehend from the comparison between the RMBPT(3)$^w$ and RMBPT(3)$^d$ results, which are propagated to all-orders in the RCC theory. To define Core contributions in the line of CPDF-RPA method, the RCC theory of $E1_{\text{PV}}$ can be derived either treating the $H_W$ and $D$ operators simultaneously as external perturbation or perturbing wave functions by considering one of these operators as external perturbation and evaluating matrix element of the other operator in the normal RCC theory framework similar to that is discussed in Ref. \cite{SahooPRL}. In order to understand the Core and Valence contributions to the $E1_{\text{PV}}$ amplitudes from our RCCSD method, we present the $E1_{\text{PV}}$ values of the $6S-7S$ and $6S-5D_{3/2}$ transitions in $^{133}$Cs from individual RCCSD terms In Table \ref{tab7}. 

Using definitions of $\Omega_0$ and $\Omega_v$ in the RCC theory and following Eqs. (\ref{eqcc}) and (\ref{eqvc}), we categorize the results into Core and Valence correlation contributions. The Core correlations arising through $\bar{D} T_1^{(1)}$ and c.c. terms contain correlation contributions from both the singly and doubly excited configurations. By analysing the RMBPT(3)$^w$ diagrams contributing to the $T_1^{(1)}$ amplitude determining equation shown in Fig. \ref{RCC-T1}, it can be understood that the $\bar{D} T_1^{(1)}$ and c.c. terms contain the Core contributions of the CPDF method, pair-correlation contributions of the RMBPT(3)$^w$ method to all-orders and many more. By analyzing diagrams from $\bar{D} T_1^{(1)}$ and their breakdown in terms of the RMBPT(3)$^w$ method carefully, it is evident that this term does not include Core contributions arising through RPA and some of the contributions that arise through the CPDF-RPA method. Similarly, all the Valence correlation contributions from the CPDF method, RPA and CPDF-RPA* method are included through the $\bar{D} S_{1i}^{(1)} + S_{1f}^{(1)\dagger} \bar{D} $ terms in the RCCSD method. They also include many contributions that can appear through the BO-correlation technique and beyond. However, a lot more correlation contributions to $E1_{\text{PV}}$ arise through other RCCSD terms among which correlation contributions arising through $\bar{D} S_{2i}^{(1)}$, $\bar{D} T_{1/2}^{(1)}S_{1/2i}^{(0)}$, $T_{1/2}^{(1)\dagger} \bar{D} S_{1/2i}^{(0)}$, such terms but replacing $S_{1/2i}^{(0/1)}$ operators with $S_{1/2f}^{(0/1)\dagger}$, $S_{1/2f}^{(0)\dagger} \bar{D} S_{1/2i}^{(1)}$, $S_{1/2f}^{(1)\dagger} \bar{D} S_{1/2i}^{(1)}$ etc. terms in the RCCSD method. Obviously, these contributions are not present in the CPDF-RPA* method and many of them cannot be considered as a part of the BO-correlation method. Moreover, corrections to the entire correlation contributions including that appear through the CPDF-RPA method due to normalization of the wave functions (given as `Norm') are quoted separately in the above table and they are found to be non-negligible. The most prominent DCP contributions are absorbed through the $\bar{D} S_{2i}^{(1)} + S_{2f}^{(1)\dagger} \bar{D} $ terms in the RCCSD method. Along with some of Core contributions from the CPDF-RPA method (like the ones appears in the RPA) are also included through these terms in the RCCSD method. In addition, non-linear terms $\bar{D} T_{1/2}^{(1)}S_{2i}^{(0)}$, $T_{2}^{(1)\dagger} \bar{D} S_{1i}^{(0)}$, $S_{2f}^{(0)\dagger} \bar{D} S_{2i}^{(1)}$ etc. including their c.c. terms posses a lot more Valence correlation contributions that are beyond the scope of considering by the combined CPDF-RPA and BO-correlation methods. 

\section{Summary}

By employing a number of relativistic many-body methods at different levels of approximation such as finite-order perturbation theory, coupled-perturbed Dirac-Fock method, random  phase approximation, combined coupled-perturbed Dirac-Fock and random phase approximation method, and relativistic coupled-cluster theory, we investigated various roles of core and valence correlation effects in the calculations of the parity violating electric dipole amplitudes of the $6S \rightarrow 7S$ and $6S \rightarrow 5D_{3/2}$ transitions in $^{133}$Cs. From this analysis, we were able to address a long standing issue of getting opposite signs to the core correlation contribution to the parity violating electric dipole amplitude of the aforementioned $6S \rightarrow 7S$ transition using the combined coupled-perturbed Dirac-Fock and random phase approximation methods. We also analysed results from the sum-over-states approach and first-principle calculations using the relativistic coupled-cluster method with singles and doubles approximation to figure out the missing contributions in the former approach. Inclusion of these missing contributions through the combined coupled-perturbed Dirac-Fock, random phase approximation and Br\"uckener-orbital correlation methods is compared with the first-principle calculations using the coupled-cluster method. This comparison shows that the first-principle approach using the relativistic coupled-cluster theory incorporates electron correlation effects due to the Dirac-Coulomb Hamiltonian more rigorously than the other methods mentioned above in the evaluation of parity violating electric dipole amplitudes in the $^{133}$Cs atom.      

\section*{Acknowledgement}

The computations reported in the present work were carried out using the Vikram-100 HPC cluster of the Physical Research Laboratory (PRL), Ahmedabad, Gujarat, India.

\section*{DATA AVAILABILITY}
The data that supports the findings of this study are available within the article.

\begin{figure}[t!]
    \centering
\includegraphics[height=44mm,width=82mm]{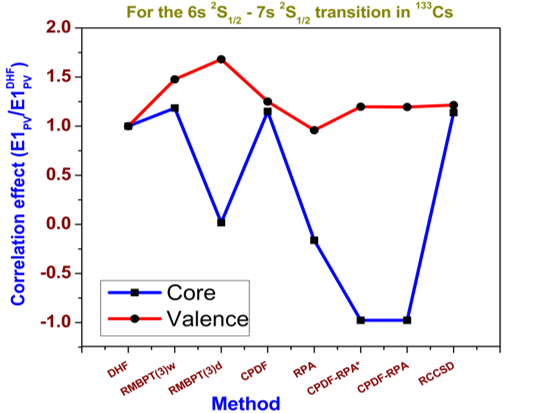}
\end{figure}
\end{document}